\documentclass[a4paper,11pt]{article}
\usepackage{graphicx}
\usepackage[USenglish,shorthands=off]{babel}
\usepackage{amsmath,amssymb,amsthm}
\usepackage{ulem}
\usepackage{bm}
\usepackage[authoryear]{natbib}
\usepackage{indentfirst}
\usepackage[labelfont=bf]{caption}
\DeclareTextCompositeCommand{\k}{LY1}{a}
  {\oalign{a\crcr\noalign{\kern-.27ex}\hidewidth\char7}}
  \usepackage{polski}
\makeatletter
\newcommand{\leqnomode}{\tagsleft@true}
\newcommand{\reqnomode}{\tagsleft@false}
\makeatother
\DeclareSymbolFont{symbolsC}{U}{pxsyc}{m}{n}
\DeclareMathSymbol{\coloneqq}{\mathrel}{symbolsC}{"42}
\theoremstyle{definition}  
\renewcommand{\bibname}{References}

\renewcommand\bibname{References}
\newcommand{\Msun}{\ensuremath{\rm \,M_\odot}}

\newcommand{\msun}{\ensuremath{\rm \,M_\odot}}

\DeclareMathAlphabet\mathbfcal{OMS}{cmsy}{b}{n}
\newcommand{\bS}{$\mathbfcal{\,S}$}

\begin{document}

\begin{figure}[h!]
    \centering
    \includegraphics[width=14cm,height=7cm]{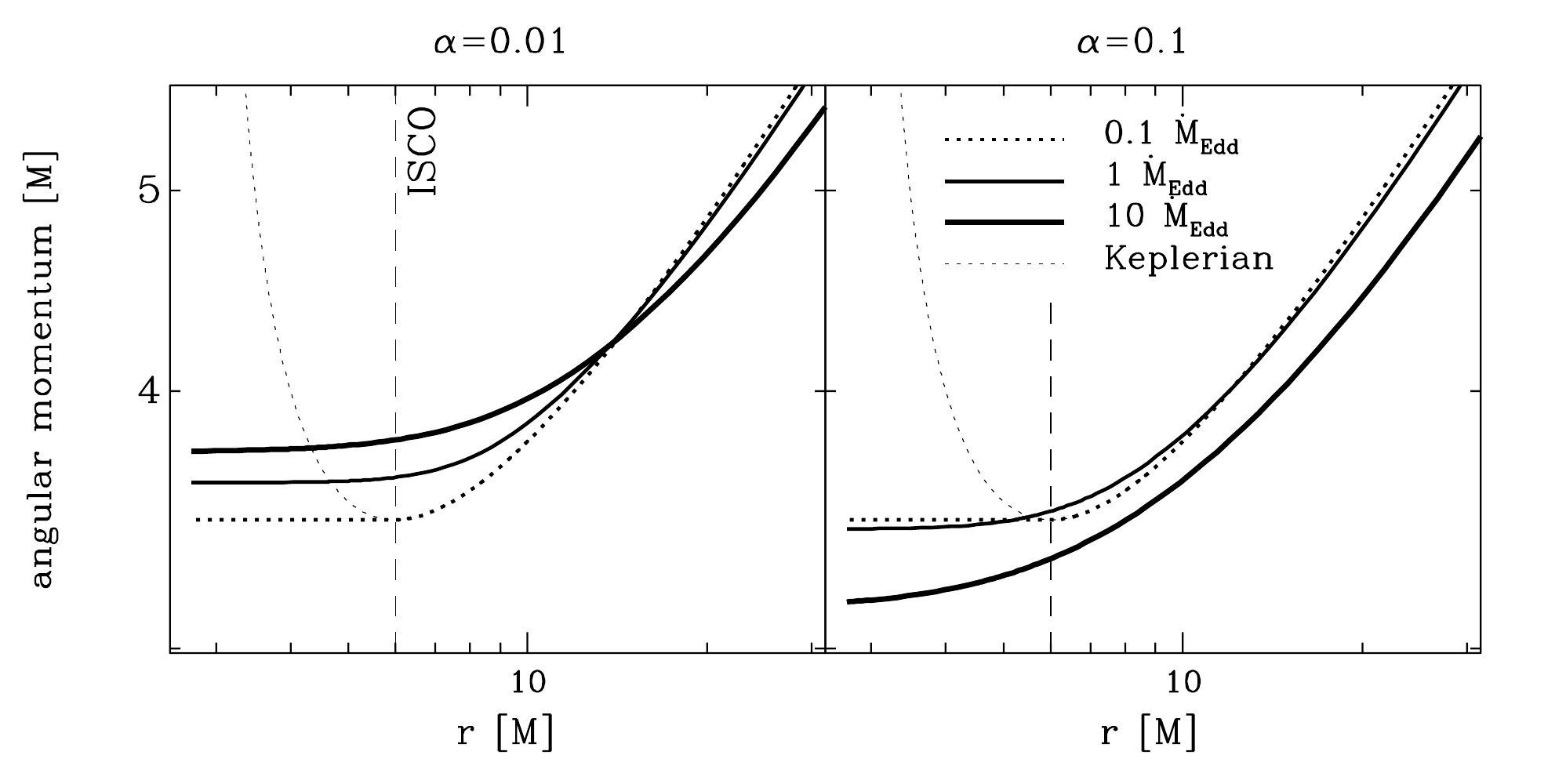}
\caption{Angular momentum profiles for slim disk solutions with
$\alpha=0.01$ (left panel) and $\alpha=0.1$ (right panel). In both panels,
three curves are presented for sub-Eddingtonian, Eddingtonian, and
super-Eddingtonian accretion rates. Since $H/R \sim \dot M(R)/\dot M_{\rm Edd}$ \citep{FKR}, only the sub-Eddingtonian curve
corresponds to a thin disc, for which clearly the torque vanishes at the inner edge: constant angular momentum below the ISCO. The thin dotted line presents the
Keplerian angular momentum profile. The angular momentum and the radius
are in units of $M$ (i.e. $R_g$, with $c=G=1$).[{\sl From \citep{Abramowicz1010}}]}.
    \label{fig:angmom}
\end{figure}

\begin{center}
%\vspace{0.5cm}

%\LARGE{\bf CHAPTER}

%\vspace{0.5cm}
{\huge\bf  AGN Accretion Discs\footnote{Updated version of Chapter~3 in {\bf Active Galactic Nuclei}; F. Combes ed.; iSTE/Wiley 2022; DOI:10.1002/9781394163724. One figure (Fig. \ref{fig:angmom}) and several references added. {\sl This chapter is a graduate--student level lecture, not a review article.} }} 

\vspace{1.0cm}

{\Large Jean-Pierre LASOTA}$^{1,2}$\\

\vspace{1.0cm}
\small{
$^1$ Nicolaus Copernicus Astronomical Center, Polish Academy of Sciences,
           Warsaw, Poland\\ 
\vspace{0.25cm}
$^2$ Institut d'Astrophysique de Paris, CNRS et Sorbonne Universit\'e,
           Paris\\
E-mail: lasota@iap.fr\\
}
\end{center}

\begin{figure}[h!]
    \centering
    \includegraphics[width=4.0cm,height=6.0cm,angle=0]{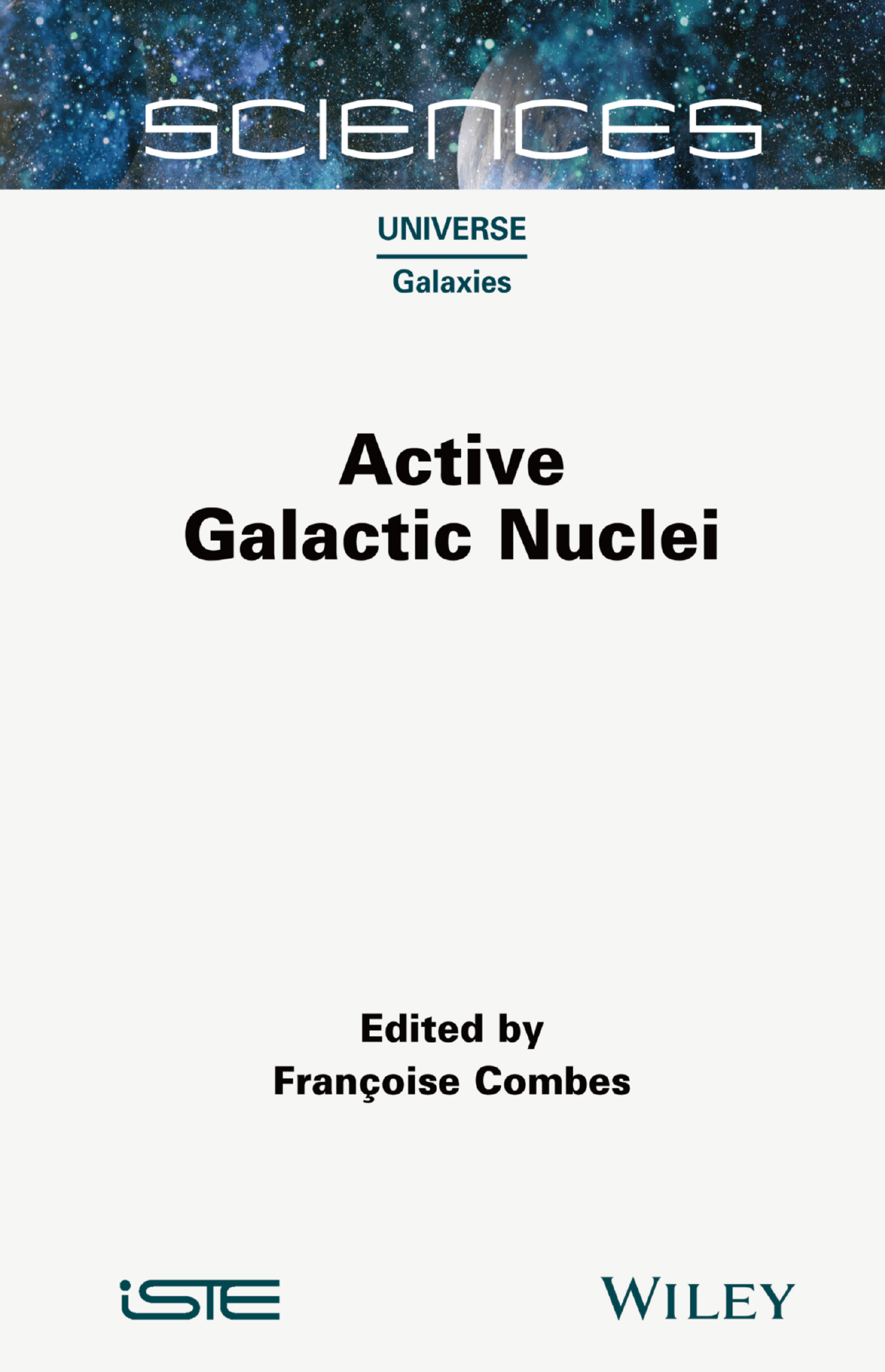}
    \caption*{}
\end{figure}

%\vspace{1.0cm}

%\begin{document}
%
%\title{}
%\author{Jean-Pierre Lasota}
%%\institute{Jean-Pierre Lasota \at 
%%Institut d'Astrophysique de Paris, CNRS et Sorbonne Universit\'e, UMR 7095, 98bis Bd Arago, 75014 Paris, France;\\
%%Nicolaus Copernicus Astronomical Center, Polish Academy of Sciences, Bartycka 18, 00-716 Warsaw, Poland; \\ 
%%\email{lasota@iap.fr}
%%}
%\date{}
%\maketitle

\section{Introduction}
\label{sec:intro}

{\sl Accretion discs}  are ubiquitous in the Universe. The spectacular 
ALMA image of the protostellar disc in HL Tau is  breathtaking and the M87* black-hole
silhouette observed by the Event Horizon Telescope (EHT) showed us the light emitted by matter as close to the event horizon as possible.

Understanding accretion discs around black holes is interesting in itself because of the fascinating and complex physics
involved but is also fundamental for understanding the coupled evolution of galaxies and their nuclear black holes, i.e.
fundamental for understanding the growth of structures in the Universe. The chance that inflows onto black holes
are strictly radial, as assumed in many models, are slim.

Accretion discs in active galactic nuclei (AGNs) play a central role in the complex engine operating around
the central supermassive black hole. They are the heart of the accretion-ejection processes, playing the role of a rugby-union
fly-half, receiving and passing matter, energy and magnetic fields to various players of the inflow-outflow team. 
Accretion disc, in their standard, geometrically-thin version, have the advantage of being the only team-members for
which a well-confirmed, even if not perfect, physical model exists. This chapter concentrates on this model, its properties and observational
consequences, as well as its possible generalisations to situations where it loses its validity.

\section{Black holes}

Black holes which are at the center of all AGN are purely general-relativistic objects and the description of
the properties and behaviour of matter and light in their vicinity requires the use of the formalism of
Einstein's theory of gravitation. Unfortunately, this theory, now used on a daily basis in astronomy
(e.g. observations of the Universe in gravitational waves) and even in everyday life (GPS in our smartphones),
has still not made it to the physics curriculum in most universities. We will therefore present here only
the most important facts about the physics in the vicinity of black holes without derivation and with minimal  justification \footnote{An excellent introduction to the general theory of relativity can be found in \citet{Hartle03}}.

\subsection{The horizon}
The Schwarzschild radius (radius of a non-rotating black hole) is
\begin{equation}
R_S=\frac{2GM}{c^2}= 2.95\times 10^{13}  \left(\frac{M}{\rm 10^8\Msun}\right)\, {\rm cm} = 9.58 \times 10^{-6} \left(\frac{M}{\rm 10^8\Msun}\right) \,  \rm pc,
\end{equation}
where $G$ is the gravitational constant, $M$ is the mass of the gravitating body and $c$ the speed of light.

Quite often in the literature and also in the present book, one finds the so-called \textsl{ gravitational radius} $R_g=0.5R_S$ used as unit of length.
Here we prefer to use as unit of length the Schwarzschild radius $R_S$.

When the black hole is rotating it is represented by the Kerr solution and the radius of its horizon\footnote{This is the radius of the external horizon. There exists also an inner horizon of no astrophysical interest.} is equal to
\begin{equation}
\label{rh}
R_H = \frac{GM}{c^2}+ \left[\left(\frac{GM}{c^2}\right)^2 -
\left(\frac{J}{Mc}\right)^2 \right]^{1/2}= \frac{GM}{c^2}\left[1 + \left(1 -
a^2 \right)^{1/2}\right],
\end{equation}
where $J$ is the black hole's angular momentum and its dimensionless angular momentum is defined as\footnote{One should be careful not to confuse this
dimensionless ``a'' with the ``a'' often used in the general-relativistic literature defined as $J/Mc$ which has dimension of length.} 
\begin{equation}
a:= \frac{Jc}{GM^2}.
\end{equation}
Therefore a horizon, i.e. a black hole, exists only for $0 \leq a \leq 1$. For $a > 1$, the Kerr solution represents a space-time
containing a naked (not covered by a horizon, i.e. observable) singularity and time-like trajectories that violate causality. The
cosmic censorship hypothesis \citep{Penrose0169}, according to which all singularities in the real Universe (except for the Big-Bang) are covered
by an event horizon remains to be proved. However, the third law of black-hole thermodynamics guarantees that no black hole
can be spun-up to $a=1$.
\begin{figure} [h!]
\begin{center}
\includegraphics[angle=0,width=0.9\textwidth]{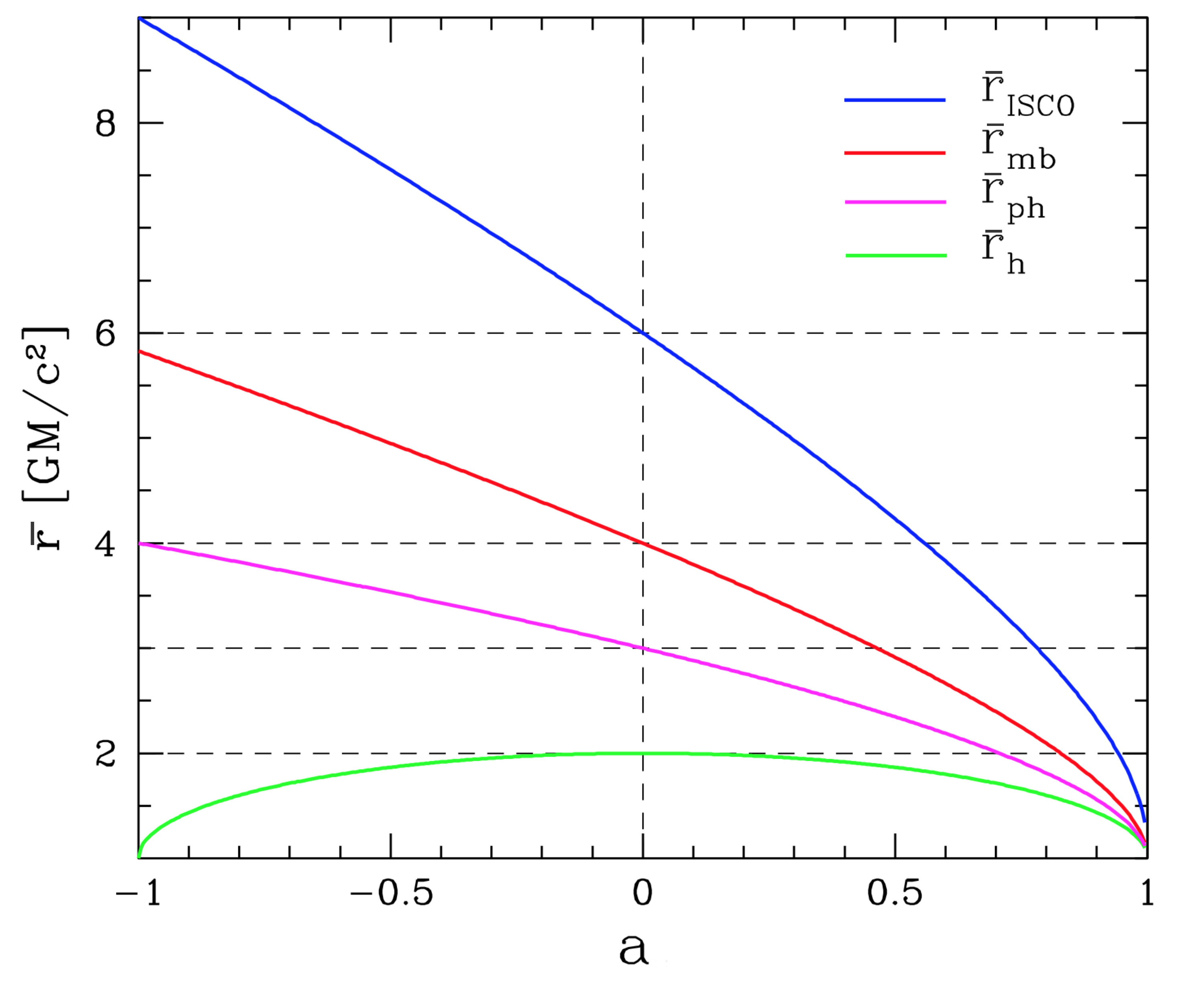}
\caption{Radii (in units of $R_g$)  of the characteristic orbits in the Kerr metric as a function of $a$. The innermost stable circular orbit: ${r}_{\rm ISCO}$, the marginally bound orbit ${r}_{\rm IBCO}$ (marked $\bar{r}_{\rm mb}$), the photon orbit: ${r}_{\rm ph}$ and the black hole horizon: ${r}_{H}$ (marked $\bar{r}_{\rm h}$) The negative
	values of $a$ correspond to orbits counterrotating with respect to the black hole. 
	({\sl Courtesy of \citealt{Sadowski11}}.)}
\label{fig:freq0}
\end{center}
\end{figure}

The numerical value of the maximum angular momentum of a black hole is
\begin{equation}
J_{\rm max}=
\frac{GM^2}{c} = 8.9 \times 10^{48} \left(\frac{M}{\msun}\right)^2
\rm g\ cm^2\ s^{-1}.
\end{equation}
which, for a solar--mass black hole, is comparable to the angular momentum of the Sun: $J_{\odot}= 1.63 \times 10^{48} \rm g\ cm^2\
s^{-1}$, or ${a}_{\odot}= 0.185$. 

For a black hole with the mass of the Galaxy ($M_G=1.5 \times 10^{11} \msun $),
the maximum angular momentum would be $2.0 \times 10^{71} \left({M_G}/{\msun}\right)^2
\rm g\ cm^2\ s^{-1}$, so the Galaxy whose angular momentum is $\sim 10^{74}\rm g\ cm^2\ s^{-1}$
would have to lose most of it, if it were to collapse into a black hole. The largest observed mass of a black hole seem to be $6.6 \times 10^{10}\, \msun$ (\citealt{Shemmer1004}).

\subsection{Characteristic orbits}

The space--time around a black hole is curved which implies that particles follow geodesics in a non-euclidean geometry. Far from the
black holes the motion along these geodesics is well approximated by newtonian orbits, but close the horizon the motion of bodies and light
is strongly affected by general-relativistic effects. Whereas in newtonian gravity, the angular momentum on Keplerian orbits decreases monotonically
with the square-root of the distance from the center, in general relativity it has a minimum at the orbit called ISCO, for ``innermost stable circular orbit''.
Closer to the horizon, free circular orbits are unstable, so that a Keplerian accretion disc (see below) has to end at the ISCO. ``Keplerian" means that
matter in the disc moves on approximately free-falling orbit, which is the case for low accretion rates. In the case of high accretion rates, the radial 
pressure gradients in the disc are no longer negligible and the inner disc's edge is pushed inwards, but cannot be closer to the horizon than the
innermost bound circular orbit, that we would like to call IBCO, but which is known under the name of  ``$R_{\rm mb}$'', ``mb'' corresponding to ``marginally bound".

\subsubsection*{ISCO}

The mathematical expression for the ISCO radius is complex and can be easily found in the literature. For the Schwarzschild solution $R_{\rm ISCO}= 3 R_S$.  For
maximally rotating black hole the radius of the counterrotating ISCO is at $R_{\rm ISCO}= 4.5 R_S$, while the radius of a corotating ISCO is formally at 
$R_{ISCO}= 0.5R_S=R_H$, but this is the result of a degeneracy for $a=1$ of the coordinate systems used.

For a Schwarzschild black hole the frequency associated with the ISCO at $R_{\rm ISCO}=3R_S$ is 
\begin{equation}
\nu_K(R_{\rm ISCO}) = 2.2 \,\left(\frac{M}{10^8 \msun}\right)^{-1}\ \rm \mu Hz.
\end{equation}
A quasi-periodic oscillation (QPO) of $\sim 27 \rm \mu Hz$ that could be associated with the ISCO frequency has been observed
in an active galaxy of type NLS1 (meaning Narrow Line Seyfert 1), whose mass is presumably $1 - 4 \times 10^6\msun$.

\subsubsection*{IBCO}

The IBCO is at radius:
\begin{equation}
R_{\rm IBCO}^{\pm}= \frac{GM}{c^2}\left[2 \mp a + 2\sqrt{1 \mp a}\right],
\end{equation}
which for a Schwarzschild black hole gives $R_{\rm IBCO}=2 R_S$. When $a=1$,  for a counterrotating orbit 
$R_{\rm IBCO}\approx 2.9 R_S$, in the corotating case, $R_{\rm IBCO}=0.5R_S$ for the reasons mentioned above.

\subsubsection*{Photon orbit}

The photon circular orbit around a black hole is also the Innermost Circular Orbit (ICO), 
and its radius is at
\begin{equation}
R_{\mathrm{phot}}^{\pm}= R_S \left(1 + \cos\left[\frac{2}{3}\cos^{-1}\left(\mp {a}\right)\right]\right).
\end{equation}
The ``photon orbit'' is at $1.5 R_S$ for a $a=0$, at $4R_S$ for $a=1$ retrograde orbit and, as all the other characteristic orbits,
at $0.5R_S$ for prograde rotation with $a=1$.

The photon circular orbit is unstable but in a suitable setting an observer could see a ``photon ring" produced by photons 
``wrapping around'' $R_{phot}$ before reaching the detector. It has been claimed that such a ring has been observed by the
EHT around the  M87 central black hole, but this is far from being certain \citep[see, e.g.][]{Gralla0719}.

\subsection{Binding energy and accretion efficiency}

The binding energy per unit mass of an orbit is (assuming a unit system where c=1),
\begin{equation}
\label{eq:bindingE}
{E}_{\rm bind}=1 - {E}_K,
\end{equation}
where ${E}_K$ is the kinetic energy per unit rest-mass of a Keplerian orbit (in the newtonian case $E_{K}= (GM/2c^2R$).

At the ISCO  the binding energy per unit rest-mass is
\begin{itemize}
\item $1 - \sqrt{8/9} \approx 0.06$ for $a=0$ 
\item $1 - \sqrt{1/3} \approx 0.42$ for $a=1$. 
\end{itemize}
Therefore this corresponds to the efficiencies
of accretion in a Keplerian disc around a black hole.

By definition the binding energy at the IBCO is equal to zero. Therefore pressure gradients in an accretion
flow that push the inner disc towards the black hole reduce accretion efficiency.

\subsection{Rotating space}

A purely general relativistic effect, with no newtonian equivalent or analogue, is the rotation of space:
the rotation of a gravitating body ``drags'' with it the surrounding space. The angular velocity of this
dragging is:
\begin{equation}
\label{eq:LTomega}
\Omega_{\mathrm{LT}}\approx \frac{2GJ}{c^2R^3},
\end{equation}
where $J$, as before is the black-hole angular momentum. ``LT'' stands for ``Lense--Thirring'' -- the names
of the two discoverers of this effect in the weak-field approximation of the Einstein's theory of gravitation.
$\Omega_{\mathrm{LT}}$ corresponds to the angular velocity with which the orbits of misaligned 
(not in the equatorial plane) test particles precess
around a black hole.

The corresponding precession time is
\begin{equation}
t_{\mathrm{LT}}\approx \frac{1}{\left|\Omega_{\mathrm{LT}}\right|}=\frac{c^{2} R^{3}}{2 G J}=\frac{2}{a}\left(\frac{R}{R_{\mathrm{S}}}\right)^{2} \frac{R}{c},
\end{equation}
which, close to the black hole can be as short as the dynamical time (At the disc inner edge of a $10^8\Msun$black hole: $R/c \approx 980$\,s), but since the precession is strongly differential, decreasing with radius,
far from the black hole it is entirely negligible: 
\begin{equation}
t_{\mathrm{LT}}\approx 7.1 \times 10^{10} a^{-1}\left(\frac{M}{10^{8} \mathrm{M}_{\odot}}\right)^{-2}\left(\frac{R}{1 \mathrm{pc}}\right)^{3} \text { yrs. }
\end{equation}
However, dissipation in a misaligned disc can lead to the alignment of its innermost regions with the black hole spin, creating a warp, through the so-called
Bardeen-Petterson effect \citep{BP75}, (see below).

\subsection{The ergoregion}

The surface of a black hole rotates with angular velocity
\begin{equation}
\Omega_{\rm H} = \frac{ac}{2R_H},
\end{equation}
where
$\Omega_{\rm H}$ is the angular velocity of the horizon, i.e. the
angular velocity of the horizon-forming light-rays with respect to
infinity. The horizon rotates.

The fact that a rotating black-hole induces rotation of space has two important astrophysical implication. First, it creates a torque 
which forces the matter in the tilted disc towards the equatorial plane in the so-called \textsl{Bardeen-Petterson effect}.
This effect, strongest near the rotating body's surface, induces the disc precession. 
Second, in the case of a Kerr black hole,
there exists an ergoregion between the horizon and the surface defined by
\begin{equation}
\label{eq:ergoreg}
R_{\rm ergo}(\theta)=\frac{GM}{c^2}\left[1 + \left(1 -
a^2 cos^2\theta\right)^{1/2}\right]
\end{equation}
where being at rest with respect to infinity is impossible: to resist the rotation of space would imply local superluminal
motion. 

This implies that the ergoregion contains trajectories with negative mass-energy as seen 
from infinity, although the energy
measured in a frame rotating with the space is positive. If e.g., an ingoing particle with energy $E_1 >0$ decays in the ergoregion into two particles, one with energy $E_2>0$, which escapes to infinity,
the other with $\Delta E_H <0$ which falls into the black hole (negative-energy particles cannot leave the 
ergoregion because this would require accelerating to speeds larger than the speed of light),
 $E_2 = E_1 - \Delta E_H > E_1$, so that the outgoing particle is more energetic than the ingoing one. 
 This energy has been gained at the expense of the rotational energy of the black hole:
 the latter absorbed negative energy and negative angular momentum, since negative-energy particles 
 counter-rotate with respect to the black-hole spin. These particles carry negative angular momentum. 
 The electromagnetic version of this \textsl{mechanical Penrose process} allows tapping the black-hole 
 rotational energy which is supposed to be the source of relativistic
jets observed in some AGN (see Sect. \ref{sec:disc-jets}).

\subsection{Eddington accretion rate}

We will define the  Eddington luminosity and accretion rate as
\begin{align}
\dot{M}_{\rm Edd} & =\frac{L_{\rm Edd}}{\eta c^2}=\frac{1}{\eta}\frac{4\pi GM}{c\kappa_{es}} \nonumber \\
& = 1.6\times 10^{26}\eta_{0.1}^{-1} \left(\frac{M}{\rm 10^8 M_{\odot}}\right)\,{\rm g\,s^{-1}} \\
& = 2.5 \, \eta_{0.1}^{-1} \left(\frac{M}{\rm 10^8 M_{\odot}}\right) \rm M_{\odot}\,\rm yr^{-1},
\end{align}
where $\eta=0.1\eta_{0.1}$ is the radiative efficiency of accretion, $\kappa_{es}$ the electron scattering (Thomson) opacity.
%\end{svgraybox}
We will often use accretion rate measured in units of Eddington accretion rate:
\begin{equation}
\label{eq:mdot}
\dot m=\frac{\dot M}{\dot{M}_{\rm Edd}}\, \ \ \ \ \ \ \ \ \ {\rm and \  use}\ \ \ \ \ \ \ \  m_{8}= \frac{M}{10^8\, \rm M_{\odot}}.
\end{equation}
One can find in the literature definitions of $\dot{M}_{\rm Edd}$ implicitly 
using $\eta= 1$, or $\eta=0.2$ so care is recommended
when using published numerical values of the accretion rate in Eddington units.

%\runinhead{Additional reading:} There are excellent general reviews of accretion disc physics, they can be found in references \cite{Blaes14}, \cite{FKR} \cite{Katoetalbook} and \cite{Spruit10}.
%%%%%%%%%%%%%%%%%%%%%%%%%%%%%%%%%%%%%%%%%%%%%%%%%%%%%%%%%%%%%%%
\section{Disc driving mechanism; viscosity}
\label{sec:viscosity}

Very early in the studies of accretion disc physics, it became clear that, if their driving mechanism
is viscous, an ``anomalous", turbulent viscosity must be at work because astrophysical discs are too large
(and their density too small) for the molecular viscosity to be efficient.
After years of uncertainty, it is now obvious that the
turbulence in ionized Keplerian discs is due to the presence of a Magneto-Rotational Instability
(MRI) also known as the Balbus-Hawley mechanism
\citep{BalbusHawley91}, occurring in \textsl{weakly} magnetised, differentially rotating plasma.
However, despite impressive developments, numerical simulations, even in their global 3D form, still suffer 
from weaknesses that, in most cases, make their direct application to real accretion flows almost infeasible. 

One of the problems is related to the value of the ratio of the (vertically averaged) total stress to thermal (vertically averaged) pressure
\begin{equation}
\label{eq:numeralpha}
\alpha=\frac{\langle{\tau_{R\varphi}}\rangle_z}{\langle P \rangle_z},
\end{equation}
where the stress $\tau_{R\varphi}$ is the sum of the Maxwell and Reynolds stresses:
\begin{equation}
\tau_{R\varphi}=- \frac{B_{R}B_{\varphi}}{4\pi} + \rho v_R \delta v_{\varphi},
\end{equation}
where indices $r$ and $\varphi$ denote, respectively the radial and azimuthal components
of the magnetic intensity $\mathbf{B}$, $\bf{v}$ and $\mathbf{\delta v}$ vectors, and $\rho$ denotes density.
$\delta v_{\varphi} = v_{\varphi} + 3\Omega_K R/2$ is the difference between the azimuthal velocity and
the mean rotational velocity in the disc, with $\Omega_K$ being the Keplerian orbital angular speed (Eq. \ref{eq: OmegaK}).
$P = P_g + P_{rad}$ is the sum of the thermal pressures; magnetic pressure is not included in this definition.
As we will see in the next section, and in the rest of this Chapter, the $\alpha$ parameter has played a crucial role in the development 
of accretion disc astrophysics. Its value ( $0\leq \alpha \leq 1$) can be determined from observations of light variability 
of some accreting systems, as will be shown below.

According to most MRI simulation $\alpha \sim 10^{-3}$ whereas observations of ionised discs around white dwarfs in binary
systems called cataclysmic variables, which are the best ``life-size" realisation of MRI discs, unambiguously show that $\alpha\approx 0.1 - 0.2$ \citep{Kotko1209}.
Disc outbursts in X-ray binaries suggest even $\alpha\approx 0.2 - 2.0$\citep{Tetarenko0218}.
The effects of convection at temperatures $\sim 10^4$\,K increase $\alpha$ to values $\sim 0.1$ and a net magnetic field
in the region might bring its value up to the observed one. On the other hand some observations of AGN variability
give values of the viscosity parameter: $0.01 \leq \alpha \leq 0.03$ for $0.01 \leq L/L_{Edd} \leq 1.0$ \citep{Starling0104}, 
still higher than the
fiducial MRI-simulation value.

Another problem is related to cold discs.
For the standard MRI to work, the degree of ionization in a weakly magnetized,
quasi-Keplerian disc must be sufficiently high to produce the instability that leads 
to a breakdown of laminar flow into turbulence;
the latter being the source of viscosity driving accretion onto the central body. 
In cold discs the ionized fraction is very small and might be
insufficient for the MRI to operate. In any case in such a disc non-ideal MHD effects are always important. All these problems
still await their solution even if a lot of progress has been made, especially in the context of protostellar accretion discs.

Finally, and very relevant to the subject of this Chapter, there is the question of stability of discs in which the pressure
is due to radiation and opacity to electron scattering. According to theory, such discs should be violently (thermally)
unstable but observations of systems presumed to be in this regime totally infirm this prediction. 

\subsection{The $\alpha$--prescription}

The $\alpha$--prescription is a rather simplistic description of the accretion disc
physics but before one is offered better and physically more reliable options its
simplicity makes it the best possible choice and has been the main source of progress
in describing accretion discs in various astrophysical contexts.

One keeps in mind that the accretion--driving viscosity is of magnetic origin, but one
uses an effective hydrodynamical description of the accretion flow.
The hydrodynamical stress tensor is 
\begin{equation}
\label{eq:stress}
\tau_{R\varphi}= \rho \nu \frac{\partial v_{\varphi}}{\partial R}=\rho \nu \frac{d\Omega}{d\ln R},
\end{equation}
where $\rho$ is the density, $\nu$ the kinematic viscosity coefficient and $v_\varphi$ the azimuthal velocity ($v_\varphi=R\Omega$).

In 1973 Shakura \& Sunyaev  proposed the (now famous) prescription
\begin{equation}
\label{eq:alphaP}
\tau_{R\varphi}=\alpha P,
\end{equation}
where $P$ is the total thermal pressure and $\alpha \leq 1$. This leads to
\begin{equation}
\label{eq:nualphaP}
\nu= \alpha c_s^2 \left[\frac{d\Omega}{d\ln R}\right]^{-1},
\end{equation}
where $c_s = \sqrt{P/\rho}$ is the isothermal sound speed and $\rho$ the density.
For the Keplerian angular velocity 
\begin{equation}
\label{eq: OmegaK}
\Omega=\Omega_K=\left(\frac{GM}{R^3}\right)^{1/2}
\end{equation}
this becomes
\begin{equation}
\label{eq:nualphaOmega}
\nu=\frac{2}{3} \alpha c_s^2/\Omega_K.
\end{equation}
Using the approximate hydrostatic equilibrium (Eq. \ref{eq:mec_approx}) one can write this as
\begin{equation}
\label{eq:nu}
\nu \approx \frac{2}{3}\alpha c_s H.
\end{equation}

Multiplying the rhs of  Eq. (\ref{eq:stress}) by the ring length ($2\pi R$) and averaging over the (total = $2H$) disc height one obtains the expression for the {\sl total torque}
\begin{equation}
\label{eq:stressH}
{\mathfrak T}= 2\pi R \Sigma \nu R\frac{d\Omega}{d\ln R},
\end{equation}
where
\begin{equation}
\label{eq:Sigma}
\Sigma=\int^{+\infty}_{-\infty}\rho\,dz .
\end{equation}
For a Keplerian disc 
\begin{equation}
\label{eq:torqueK}
{\mathfrak T}= 3\pi \Sigma \nu \ell_K,
\end{equation}
($\ell_K=R^2\Omega_K$ is the Keplerian \textsl{specific}(i.e. per unit mass) angular momentum.)

The viscous heating is proportional to $\tau_{r\varphi}(d\Omega/dR)$.
In particular the viscous heating rate per unit volume is
\begin{equation}
\label{eq:visheat1}
q^+= -\tau_{r\varphi}\frac{d\Omega}{d\ln R},
\end{equation}
which for a Keplerian disc, using Eq. (\ref{eq:alphaP}), can be written as
\begin{equation}
\label{eq:qplusalphaP}
q^+=\frac{3}{2}\alpha \Omega_K P,
\end{equation}
and  the viscous heating rate per unit surface is therefore
\begin{equation}
\label{eq:vischeat2}
Q^+=\frac{{\mathfrak T\Omega^{\prime}}}{4\pi R}=\frac{9}{8} \Sigma \nu \Omega_K^2.
\end{equation}
(The denominator in the first rhs is $2\times 2\pi R$ taking into account the existence of two disc surfaces.)

%\runinhead{Additional reading:} References \cite{Balbus11}, \cite{BalbusHawley91} and \cite{Hirose14}.

%%%%%%%%%%%%%%%%%%%%%%%%%%%%%%%%%%%%%%%%%%%%%%%%%%%%%%%%%%%%%%%%

\section{Geometrically thin Keplerian discs}
\label{sec:thin}

The 2D structure of geometrically thin, non--self-gravitating, axially symmetric accretion discs 
can be split into a 1D+1D structure
corresponding to a hydrostatic vertical configuration and radial quasi-Keplerian viscous flow. These two 1D structures
are coupled through the viscosity mechanism transporting angular momentum and providing the local release
of gravitational energy.

\subsection{Disc vertical structure}
\label{subsec:vstruct.1}

The vertical structure can be treated as a one--dimensional star with two essential differences:
\begin{enumerate}
\item the energy sources are distributed over the whole height of the disc, while in a star they are limited to the nucleus,
\item the gravitational acceleration \textsl{increases} with height because it is given by the tidal gravity of the accretor, while in stars the (self)gravity
decreases as the inverse square of the distance from the center.
\end{enumerate}

Taking these differences into account the standard stellar structure equations adapted to the description of the disc vertical structure are listed
below.

\begin{itemize}
\item Hydrostatic equilibrium

The gravity force is counteracted by the force produced by the pressure gradient:
\begin{equation}
\label{eq:vertichydro}
\frac{dP}{dz}=\rho g_z,
\end{equation}
where $g_z$ is the vertical component (tidal) of the accreting body gravitational acceleration: 
\begin{equation}
\label{eq:vertacc}
g_z=\frac{\partial}{\partial z }\left[\frac{GM}{(R^2 + z^2)^{1/2}}\right]\approx \frac{GM}{R^2}\frac{z}{R}.
\end{equation}
The second equality follows from the assumption that $z\ll R$.
Denoting the typical (pressure or density) scale-height by $H$ the condition of geometrical thinness of the disc is $H/R\ll 1$
and writing $dP/dz \sim P/H$, Eq. (\ref{eq:vertichydro})
can be written as
\begin{equation}
\label{eq:mec_approx}
\frac{H}{R}\approx \frac{c_s}{v_K},
\end{equation}
where  $v_K=\sqrt{GM/R}$ is the Keplerian velocity and we made use of Eq. (\ref{eq:vertacc}). 
From Eq. (\ref{eq:mec_approx}) it follows that
\begin{equation}
\label{eq:dyntime}
\frac{H}{c_s} \approx \frac{1}{\Omega_K}= :t_{\rm dyn},
\end{equation}
where $t_{\rm dyn}$ is the dynamical time.\\
For the parameters of interest
\begin{equation}
\label{eq:dynparam}
t_{\rm dyn}= 1.4 \times 10^3 m_8 r^{3/2} \,\rm s.
\end{equation}
Here,  and in what follows $r :=R/R_S$, and $m_8 :=$ M/$10^8\Msun$.
Therefore, e.g., for $R\approx 10^{16}$cm (or 330 $R_S$),
and a $10^8\Msun$ black hole, the dynamical time is about $3.3$ months.

\item Mass conservation

In 1D hydrostatic equilibrium the mass conservation equation takes the simple form of
\begin{equation}
\label{eq:mass_cons}
\frac{d \varsigma}{dz} = 2 \rho,
\end{equation}
where $\varsigma$ is the surface
density between $-z$ and $+z$.
\medskip

\item Energy transfer - temperature gradient

\begin{equation}
\label{eq:nabla}
\frac{d\ln T}{dz} = \nabla \frac{d \ln P}{dz}.
\end{equation}
For radiative energy transport
\begin{equation}
\label{eq:nablarad}
\nabla_{\rm rad} = \cfrac{}{}{\kappa_{\rm R} P F_{\rm z}}{4 P_r c g_{\rm z}},
\end{equation}
where $P_r$ is the radiation pressure and $\kappa_{\rm R}$ the Rosseland mean opacity.
From Eqs. (\ref{eq:nabla}) and (\ref{eq:nablarad}) one recovers the familiar expression for
the radiative flux
\begin{equation}
\label{eq:radflux}
F_z= - \frac{16}{3}\frac{\sigma T^3}{\kappa_{\rm R}\rho}\frac{\partial T}{\partial z}=- \frac{4\sigma}{3\kappa_{\rm R}\rho}\frac{\partial T^4}{\partial z}
\end{equation}
($F_z$ is positive because the temperature decreases with $z$ so ${\partial T}/{\partial z}<0$).

The photosphere is at optical thickness $\tau\simeq 2/3$ (see Eq. \ref{t2}).
The boundary conditions are: $z = 0$, $F_z = 0$, $T = T_c$, $\varsigma = 0$ at
the disc midplane;  at the disc photosphere $\varsigma=\Sigma$ and
$T^4(\tau=2/3) = T^4_{\rm eff}$. For a detailed discussion of radiative transfer, temperature stratification and boundary conditions see Sect. \ref{subsect:rad}.

In the same spirit as Eq. (\ref{eq:mec_approx}) one can write Eq. (\ref{eq:radflux}) as 
\begin{equation}
\label{eq:rad_approxx}
F_z\approx  \frac{4}{3}\frac{\sigma T_c^4}{\kappa_{\rm R}\rho H}= \frac{8}{3}\frac{\sigma T_c^4}{\kappa_{\rm R}\Sigma},
\end{equation}
where $T_c$ is the mid-plane (``central") disk temperature. Using the optical depth $\tau= \kappa_{\rm R}\rho H= (1/2) \kappa_{\rm R}\Sigma$, this can be written as
\begin{equation}
\label{eq:rad_approx2}
F_z(H) \approx  \frac{8}{3}\frac{\sigma T_c^4}{\tau}= Q^-,
\end{equation}
(see Eq. \ref{diff2} for a rigorous derivation of this formula).

In the case of convective energy transport $\nabla=\nabla_{\rm conv}$. Because convection in discs is still not well understood  there is no obvious choice for $\nabla_{\rm conv}$ so in practice prescriptions based on 
stellar physics, such as the mixing-length approximation, are used, even if it is not obvious that
they apply to accretion discs. Some evidence provided by MRI simulations suggests that they do not.

\item Energy conservation

Vertical energy conservation should have the form
\begin{equation}
\label{eq:strc1}
\frac{dF_{\rm z}}{dz } = q^+ (z),
\end{equation}
where $q^+ (z)$ corresponds to viscous energy dissipation per unit volume.
Note that, in contrast with accretion discs, stellar envelopes are in \textsl{radiative equilibrium} ${dF_{\rm z}}/{dz }=0$

The $\alpha$ prescription does not allow deducing the viscous dissipation stratification ($z$ dependence), it just says that the vertically averaged viscous torque
is proportional to pressure. Most often one assumes therefore that 
\begin{equation}
\label{eq:strc2}
q^+ (z) = \frac{3}{2} \alpha \Omega_{\rm K} P (z),
\end{equation}
by analogy with Eq. (\ref{eq:qplusalphaP}) but such an assumption is chosen because of its simplicity and not because of some physical motivation. In fact MRI
numerical simulations suggest that dissipation is not stratified in the same way as pressure.

\item The vertical structure equations have to be completed by the equation of state (EOS):
\begin{equation}
P = P_r + P_g= \frac{4\sigma }{3c}T^4 + \frac{{\cal R}}{\mu} \rho T ,
\label{eq:EOS}
\end{equation}
where ${\cal R}$ is the gas constant and $\mu$ the mean molecular weight, 
and an equation describing the mean opacity dependence on density and temperature.

\end{itemize}

\subsection{Disc radial structure}
\label{subsec:radial1}

\begin{itemize}
\item 
Continuity (mass conservation) equation has the form
\begin{equation}
\frac{\partial \Sigma}{\partial t} = - \frac{1}{R} \frac{\partial}{\partial
R} (R \Sigma v_{\rm r}) + \frac{S(R,t)}{2 \pi R}, 
\label{eq:consm}
\end{equation}
where $S(R,t)$ is the matter source (sink) term. 

\item Angular momentum conservation 

\begin{equation}
\frac{\partial \Sigma \ell}{\partial t} = - \frac{1}{ R} \frac{\partial}{\partial
R} (R \Sigma \ell v_{\rm r}) + \frac{1}{ R} \frac{\partial}{\partial R}
\left(R^3 \Sigma \nu \frac{d\Omega}{dR} \right) + \frac{S_{\ell}(R,t)}{2\pi R}.
\label{eq:consj}
\end{equation}
$\ell$ is the specific  (per unit mass) angular momentum. This conservation equation reflects the fact that angular momentum is transported through the disc by a viscous stress $\tau_{r\varphi}=R \Sigma \nu {d\Omega}/{dR}$.
Therefore, if the disc is not considered infinite (recommended in application to real processes and systems) there must be somewhere a sink of this transported angular momentum $S_\ell (R,t)$. 
\end{itemize}

For semi-detached binary systems there is both a source (angular momentum brought in by the mass transfer from 
the stellar companion) and a sink (tidal interaction taking angular momentum back to the orbit). 
In the case of accretion discs in AGN, neither the processes through which matter is fed to the disc, 
nor the mechanism removing angular momentum are well established so the form of both $S(R,t)$ and
${S_{\ell}(R,t)}$ are unknown. One can only safely assume that the equations used 
in this Section will not apply beyond the radius at which the disc becomes self-gravitating 
(see Sect. \ref{subsec:sg}) which defines a natural outer boundary of the disc considered here.

Assuming $\Omega=\Omega_K$, from Eqs. (\ref{eq:consm}) and (\ref{eq:consj}) one can obtain a diffusion equation for the surface density $\Sigma$:
\begin{equation}
\label{eq:eqdiff}
\frac{\partial \Sigma}{\partial t}=\frac{3}{R}\frac{\partial}{\partial R}\left\{ R^{1/2}  \frac{\partial}{\partial R}\left[ \nu \Sigma R^{1/2}\right]\right\}.
\end{equation}
Comparing with Eqs. (\ref{eq:consm}) one sees that the radial velocity induced by the viscous torque is
\begin{equation}
v_r= - \frac{3}{\Sigma R^{1/2}}\frac{\partial}{\partial R}\left[ \nu \Sigma R^{1/2}\right],
\label{eq:vr}
\end{equation}
which is an example of the general relation
\begin{equation}
\label{eq:vvisc}
v_{\rm visc}\sim \frac{\nu}{R}.
\end{equation}

Using Eq.(\ref{eq:nu}) one can write
\begin{equation}
\label{eq:vistime1}
t_{\rm vis} := \frac{R}{v_{\rm visc}}\approx \frac{R^2}{\nu}\approx\alpha^{-1}\frac{H}{c_s}\left(\frac{H}{R}\right)^{-2}.
\end{equation}

The relation between the viscous and the dynamical times is
\begin{equation}
t_{\rm vis}\approx  \alpha^{-1}\left(\frac{H}{R}\right)^{-2}\, t_{\rm dyn}.
\label{eq:vistime2}
\end{equation}
In thin ($H/R \ll 1$) accretion discs the viscous time is much longer that the dynamical time. In other
words, during viscous processes the vertical disc structure can be considered to be in hydrostatic equilibrium.

\begin{itemize}
\item Energy conservation\\

The general form of energy conservation (thermal) equation can be written as:
\begin{equation}
\label{eq:energy}
\rho T\frac{d s}{dR}:=\rho T\left(\frac{\partial s}{\partial t} +v_r\frac{\partial s}{\partial R}\right)= q^+ - q^- + \widetilde q,
\end{equation}
where $s$ is the entropy density, $q^+$ and $q^-$ are respectively the viscous and radiative energy density, and
$\widetilde q$ is the density of external and/or radially transported energy densities.
The term
\begin{equation}
\label{eq:adv1}
\rho T v_r\frac{\partial s}{\partial R} =: q^{\rm adv}
\end{equation}
describes radial advection of energy.

Using the first law of thermodynamics
$Tds=dU +PdV$
one can write
\begin{equation}
\label{eq:firstlaw}
\rho T\frac{ds}{dt}=\rho\frac{d U}{d t} + P\frac{\partial v_r}{\partial r},
\end{equation}
where $U={{\Re} T_{\rm c} /\mu (\gamma - 1)}$.

Vertically averaging, but taking $T=T_c$, using Eq. (\ref{eq:consm}) and assuming a gas-pressure dominated disc ($P=P_g$),
one obtains
\begin{equation}
\frac{\partial T_{\rm c}}{\partial t} +v_{\rm r} \frac{\partial T_{\rm c}}{\partial R}+
\frac{\Re T_{\rm c}}{\mu c_P} \frac{1}{R} \frac{\partial (R v_{\rm r})}{\partial R} 
= 2\frac{Q^+ -Q^-}{c_P \Sigma} + \frac{\widetilde Q}{c_P\Sigma} ,
\label{eq:heat}
\end{equation}
where $Q^+$ and $Q^-$ are respectively the heating and cooling rates per unit surface.
${\widetilde Q}=Q_{\rm out} + J$ with $Q_{\rm out}$ corresponding to energy contributions 
by the mass-transfer stream and tidal torques; $J(T,\Sigma)$ represent radial energy fluxes
that are a, more or less, ad hoc addition to the 1D+1D scheme to which they do not belong: indeed
such a scheme assumes that radial gradients ($\partial/\partial R$) of physical quantities can be neglected
when compared with gradients in the vertical direction. That is also the reason that as long as
$H/R \ll 1$, $q^{\rm adv}$ is negligible compared to $q^+$ and $q^{-}$.
\end{itemize}

The viscous heating rate per unit surface can be written as (see Eq. \ref{eq:vischeat2})
\begin{equation}
Q^+=\frac{9}{8} \nu \Sigma \Omega_{\rm K}^2
\label{eq:qplus}
\end{equation}
while the cooling rate over unit surface (the radiative flux) is obviously
\begin{equation}
Q^- =\sigma T_{\rm eff}^4.
\label{eq:qminus}
\end{equation}
In thermal equilibrium one has
\begin{equation}
\label{eq:thermal_equi}
Q^+=Q^-.
\end{equation}

The cooling time can be easily estimated from Eq. (\ref{eq:thermal_equi}).
The energy density to be radiated away is $\rho U \sim \rho c_s^2$,
so the energy per unit surface is $\sim \Sigma c_s^2$ and the cooling (thermal) time is
\begin{equation}
\label{eq:thermal_time}
t_{\rm th}=\frac{\Sigma c_s^2}{Q^-}=\frac{\Sigma c_s^2}{Q^+}\sim \alpha^{-1}\Omega_K^{-1}= \alpha^{-1}t_{\rm dyn}.
\end{equation}
Since $\alpha < 1$, $t_{\rm th}>t_{\rm dyn}$  and during thermal processes the disc can be assumed to be in (vertical)
hydrostatic equilibrium.

For geometrical thin ($H/R\ll 1$) accretion discs one has the following hierarchy of the characteristic times
\begin{equation}
\label{eq:timeshare}
t_{\rm dyn} < t_{\rm th} \ll t_{\rm vis}.
\end{equation}

(This hierarchy is similar to that of characteristic times in stars where the dynamical time 
is shorter than the thermal (Kelvin-Helmholtz) time, and the
thermal time is much shorter than the thermonuclear time-scale.)

\subsection{Self-gravity}
\label{subsec:sg}

In this Chapter we are interested in discs that are not self-gravitating, i.e. in discs where the vertical hydrostatic equilibrium is maintained against the pull
of the accreting body's tidal gravity whereas the disc's self-gravity can be neglected. We will see now under what conditions this assumption
is satisfied.

The equation of vertical hydrostatic equilibrium can be written as 
\begin{equation}
\frac{1}{\rho}\frac{dP}{dz}= - g=\left(-g_z - g_s \right)= - g_z\left(1+\frac{g_s}{g_z}\right)=:- g_z \left(1 + A\right),
\end{equation}
therefore self-gravity is negligible when $A \ll 1$.
Treating the disc as an infinite uniform plane (i.e. assuming the surface density does not vary too much with radius) one can write its self gravity as
$g_s=2\pi G\Sigma$, whereas the $z$-component of the gravity provided by the central body is $g_z=\Omega_K^2\,z$ (Eq. \ref{eq:vertacc}). Therefore evaluating $A$ at $z=H$
one gets
\begin{equation}
A_H:= \frac{g_s}{g_z}\bigg|_H = \frac{2\pi G \Sigma}{\Omega_K^2 H}.
\end{equation}
$A_H$ is related to the so-called Toomre parameter \citep{Toomre64} 
\begin{equation}
\label{eq:toomre}
Q_T:=\frac{c_s\Omega}{\pi G\Sigma},
\end{equation}
widely used in the studies of gravitational stability of rotating systems, through $A_H \approx Q_T^{-1}$.
We will therefore express the condition of negligible self-gravity (gravitational stability) as
\begin{equation}
\label{eq:nosg}
Q_T > 1.
\end{equation}

Notice that the condition of negligible disc self-gravity (Eq. \ref{eq:nosg}) is equivalent to $M_D < (H/R)\,M_{\rm BH}$ and not to just $M_D < M_{\rm BH}$,
as sometimes claimed, where $M_D$ is the disc mass, and $M_{\rm BH}$ is the mass of the accreting body, a black hole in our case.

Using Eqs. (\ref{eq:mec_approx}), (\ref{eq:nu}) and (\ref{eq:amintK}) one can write the Toomre parameter as
\begin{equation}
Q_T=\frac{3\alpha c_s^3}{G\dot M},
\end{equation}
or as function of the mid-plane 
temperature $T=10^4\,T_4$K
\begin{equation}
\label{eq:Qdisc}
Q_T\approx 0.5 \,\frac{\alpha\, T^{3/2}_4}{m_8\,\dot m}.
\end{equation}
(where $\dot M= \dot m$ M$_\odot$/yr is the accretion rate, expressed in 
solar mass per year, see below).
This shows that hot ionized ($T\gtrsim 10^4$K) discs become self-gravitating for high accretor masses and
high accretion rates. Discs in close binary systems ($m_8 \lesssim 30$) are never self-gravitating for realistic accretion
rates ($\dot m < 1000$, say) and even in intermediate-mass black holes (IMBH) binaries 
(if they exist) (hot) discs would also be free of the gravitational instability.
Around a supermassive black hole, however, discs can become self-gravitating quite close to the black hole. 
For example when
the black hole mass is $m_8=1$ a hot disc will become self-gravitating at $r \approx 100$, 
for $\dot m \sim 10^{-2}$.
In general, geometrically thin, non--self-gravitating accretion discs around supermassive black holes 
have a very limited radial extent. 
From the frequently used relation (see Sect. \ref{subsec:SS})
\begin{equation}
Q_T \approx 10^4 \alpha^{7/10}m_8^{-13/10}\dot m^{-11/20}r^{-9/8},
\label{eq:selfgagn}
\end{equation}
one obtains for the radius of self-gravitating radius
\begin{equation}
r_{\rm sg} < 4.0 \times 10^3 \alpha^{28/45}m_8^{-52/45}\dot m^{-22/45},
\label{eq:rselfg}
\end{equation}
or, e.g., for $m_8=1$ and $\dot m=0.01$, $R < 0.4$ pc.
The value of the self-gravitating radius plays a fundamental role in determining the maximum mass of a luminous (through accretion of matter) black hole \citep{King0216}.

%\runinhead{Additional reading:} References \cite{CollinZahn08}, \cite{Gammie01},\cite{Goodman03}, \cite{LinPringle87}, \cite{Paczynski78} and \cite{Toomre64}.

\subsection{Stationary discs}
\label{subsec:stationary}

In the case of stationary ($\partial/\partial t=0$) discs Eq. (\ref{eq:consm}) can be easily integrated giving
\begin{equation}
\label{eq:accrrate}
\dot M:=2\pi R \Sigma v_{\rm r},
\end{equation}
where the integration constant $\dot M$ (mass/time) is the \textsl{accretion rate}.

Also the angular momentum equation  (\ref{eq:consj}) can be integrated to give
\begin{equation}
\label{eq:amint1}
-2\pi R\Sigma v_r\ell  + 2\pi R^3\Sigma\nu\frac{d\Omega}{dr}=const.
\end{equation}
Or, using Eq. (\ref{eq:accrrate}),
\begin{equation}
\label{eq:amint2}
-\dot M \ell   + {\mathfrak T}={const.},
\end{equation}
where the torque 
\begin{equation}
\label{eq:torque}
{\mathfrak T}:= 2\pi R^3\Sigma\nu d\Omega/dr; 
\end{equation}
(for a Keplerian disc ${\mathfrak T}= 3\pi R^2\Sigma\nu\Omega_K$).

Assuming that the torque vanishes at the inner disc radius, one gets
$const. =- \dot M \ell_{\rm in}$, where $\ell_{\rm in}$ is the specific angular momentum at the
disc inner edge.
Therefore
\begin{equation}
\label{eq:amintt}
\dot M (\ell -\ell_{\rm in}) ={\mathfrak T}
\end{equation}
which is a simple expression of angular momentum conservation.

For Keplerian discs one obtains an important relation between viscosity and accretion rate
\begin{equation}
\label{eq:amintK}
\nu \Sigma =\frac{\dot M }{3\pi}\left[1 -\left(\frac{R_{\rm in}}{R}\right)^{1/2} \right].
\end{equation}
From Eqs. (\ref{eq:amintK}), (\ref{eq:qplus}),  (\ref{eq:qminus}), and the thermal equilibrium equation (\ref{eq:thermal_equi})
it follows that
\begin{equation}
\label{eq:teff}
\sigma T_{\rm eff}^4 =\frac{3}{8\pi}\frac{GM\dot M}{R^3}\left[1 -\left(\frac{R_{\rm in}}{R}\right)^{1/2} \right].
\end{equation}

This assumes only a Keplerian disc in thermal ($Q^+=Q^-$) and viscous ($\dot M=const.$) equilibrium. Since the disc is in thermal equilibrium, the emitted radiation flux cannot contain information about the heating mechanism which explains why the viscosity coefficient is absent from Eq. (\ref{eq:teff}). Steady discs do not provide information about the viscosity operating in discs or the viscosity parameter $\alpha$. To get this information one must consider (and observe) time-dependent states of accretion discs. It is often incorrectly claimed that the $R^{-3/4}$ temperature profile is a signature of the Shakura-Sunyaev solution, but obviously this profile is much more general and independent of the viscosity prescription assumed to obtain this celebrated solution (see below).

Eq. (\ref{eq:teff}) determines indeed a universal radial temperature profile for \textsl{stationary Keplerian accretion discs}
\begin{equation}
\label{eq:temprofile}
T_{\rm eff}\sim R^{-3/4}.
\end{equation}

For an optically thick disc the temperature relations $T\sim T_{\rm eff}$ and $T\sim R^{-3/4}$ should be observed if
stationary, optically thick Keplerian discs do exist in the Universe.  And vice versa, if they are observed, this proves that
such discs exist not only on paper. The $R^{-3/4}$ disc--temperature profiles have been clearly observed in bright, eclipsing cataclysmic variables, 
it seems, however, that in AGN discs these profiles are a bit steeper, as will be discussed below.

The temperature profile of stationary keplerian accretion disc is given therefore by
\begin{equation}
\label{eq:Teff_value}
T_{\rm eff}=T_{\rm in} \left(\frac{r}{3}\right)^{-3/4},
\end{equation}
where
\begin{equation}
T_{\rm in}=\left(\frac{3GM\dot M}{8\pi \sigma (3R_S)^3}\right)^{1/4} \approx 3.0 \times 10^5\,  m_{8}^{-1/2} \dot m^{1/4}\rm K,
\label{eq:Teff_in}
\end{equation}
where we assumed that $R_{\rm in}=3R_S$, i.e. the ISCO for a non-rotating black hole.
Therefore the maximum effective disc temperature $T_{\rm eff}^{\rm {max}}=0.488\, T_{\rm in}$ is located at $r =49/12 \approx 4.1$ and
in AGN corresponds typically to UV radiation.

\subsubsection{The ``no-torque condition"}

There has been a lot of discussion about the inner boundary condition
in an accretion disc around a black hole. The usual reasoning is that for a thin disc the
inner boundary is at ISCO and since circular orbits end there,
the boundary condition should be simply that the ``viscous" torque
vanishes there (there is no orbit below the ISCO to interact with). 
In fact, for geometrically thin accretion discs, the no-torque condition is
a simple consequence of conservation of angular momentum \citep{Paczynski0400}.
Numerical simulations of thin accretion discs that do not satisfy this condition
do not conserve angular momentum.

In general, if one does not assume that the torque vanishes at the inner disc edge, Eq. (\ref{eq:amint2})
will take the form
\begin{equation}
\label{eq:amint3}
\dot M (\ell -\ell_{H}) ={\mathfrak T},
\end{equation}
where $l_H$ is the specific angular momentum of the accretion flow at the black-hole surface: torques
must vanish on the horizon, since the horizon is an hypersurface causally detached from the rest of the Universe.

From Eqs. (\ref{eq:amint3}), (\ref{eq:torque}) and the viscosity prescription  $\nu \approx \alpha H^2 \Omega$,
one can obtain
\begin{equation}
v_r \approx \alpha ~ H^2 ~ \frac{{\ell}}{{\ell} - {\ell}_H} ~ \frac{d \Omega}{dr}
\approx \alpha ~ H^2 ~ \frac{{\ell}}{{\ell} - {\ell}_H} ~ \frac{ \Omega}{ r }
\approx \alpha ~ v_{\varphi} \left( \frenchspacing{H}{R} \right) ^2 ~ \frac{{\ell}}{{\ell} - {\ell}_H},
\label{vr2}
\end{equation}
where $v_{\varphi}= R \Omega$.
Equation (\ref{vr2}) does not assume that the
radial velocity is small, i.e. this equation holds within the disk as well as
within the stream below the ISCO.  
Far out in the disk, where $ {\ell} \gg {\ell}_H$, one
obtains the standard formula (see Eq. \ref{eq:vvisc})
\begin{equation}
v_r \approx \alpha ~ v_{\varphi} \left( \frac{ H}{R } \right) ^2 ,
\hskip 1.0cm R \gg R_{in} .
\end{equation}

The flow crosses the black hole surface at the speed of light and since it is subsonic in the
disc it must somewhere become transonic, i.e. go through a sonic point, which has been shown to be close to
the disc's inner edge.

At the sonic point $ v_r = c_s \approx (H/R)v_{\varphi} $, and
equation (\ref{vr2}) becomes:
\begin{equation}
\frac{v_r }{ c_s} = 1 \approx \alpha ~ \frac{H_{\rm in}}{ R_{\rm in} } ~
\frac{{\ell}_{\rm in}}{{\ell}_{\rm in} - {\ell}_H},
\hskip 1.0cm R = R_{\rm in}.
\label{bep1}
\end{equation}
If the disc is thin, i.e. $ H_{\rm in} / R_{\rm in} \ll 1 $, and the
viscosity is small, i.e.  $ \alpha \ll 1 $, then Eq. (\ref{bep1}) implies that
$({\ell}_{\rm in} - {\ell}_H)/{{\ell}_{\rm in}} \ll 1 $,
i.e. the specific angular momentum at the sonic point is almost
equal to the asymptotic angular momentum at the horizon.

In a steady state disc the torque ${\mathfrak T}$ has to satisfy the equation
of angular momentum conservation (\ref{eq:amint2}), which can be written as
\begin{equation}
{\mathfrak T}= \dot M \left( {\ell} -  {\ell}_H \right) ,
\hskip 1.0cm
{\mathfrak T}_{in} = \dot M
\left({\ell}_{in} -  {\ell}_H \right) .
\label{torquebp}
\end{equation}
%\begin{svgraybox}
Thus it is clear that for a thin, low viscosity disk the `no torque inner
boundary condition' (${\mathfrak T}_{in}\approx 0$) is an excellent approximation
\textit{following from angular momentum conservation}.
%\end{svgraybox}
However, if the
disk and the stream are thick, i.e.  $ H/r \lesssim 1 $, and the
viscosity is high, i.e. $ \alpha \sim 1 $, then the angular momentum can
vary also in the stream in accordance with the simple reasoning presented
above. In such a case the no--stress condition  at the disc inner edge might be
unsatisfied.

\subsubsection{Total luminosity and spectrum}
\label{susect:totalL}

The total luminosity of a stationary, geometrically thin accretion disc, i.e. the sum of luminosities of its two surfaces, is
\begin{equation}
\label{eq:lum1}
2\int_{R_{\rm in}}^{R_{\rm out}}\sigma T_{\rm eff}^4\,2\pi RdR=  \frac{3GM\dot M}{2}\int_{R_{\rm in}}^{R_{\rm out}}\left[1 -\left(\frac{R_{\rm in}}{R}\right)^{1/2} \right]\frac{dR}{R^2}.
\end{equation}
For $R_{\rm out} \rightarrow \infty$ this become
\begin{equation}
\label{eq:lum2}
L_{\rm disc}=\frac{1}{2}\frac{GM\dot M}{R_{\rm in}}=\frac{1}{2}L_{\rm acc}.
\end{equation}
In the disc the radiating particles move on Keplerian orbits hence they retain half of the potential energy.
When the accreting body is a black hole this leftover energy will be lost. In this case, however, the non-relativistic formula of Eq. (\ref{eq:lum2})
does not apply -- see Eq. (\ref{eq:bindingE}). 

For an optically thick disc one can assume that each of its rings radiates as black-body emitter:
\begin{equation}
I_\nu=B_\nu\left[T_{\rm eff}(R)\right]=\frac{2h\nu^3}{c^2(e^{h\nu/kT(R)}-1)}\,\left(\mathrm{erg\,s^{-1}\,cm^{-2}\,Hz^{-1}\,sr^{-1}}\right).
\label{eq:bbdisc}
\end{equation}
This is a rather crude but useful approximation. One should keep in mind, however, that it does not represent real observed disc spectra. Stars are very
optically thick but their spectra do not look like black bodies, because the observed light had to 
cross the stellar atmosphere on its way to the detector.

With such an assumption the flux at a given frequency, detected by an observer at a distance $D$ is equal to
\begin{equation}
F_\nu=\frac{2\pi \cos i}{D^2}\int^{R_{\rm out}}_{3R_S}I_\nu R\,dR = \frac{4\pi h \nu^3\,\cos i}{c^2D^2}\int^{R_{\rm out}}_{3R_S}\frac{ R\,dR}{e^{h\nu/kT(R)}-1},
\label{eq:flux_disc}
\end{equation}
where $i$ is the disc inclination.

For frequencies
\begin{equation}
\frac{kT({R_{\rm out}})}{h} \ll \nu \ll \frac{kT_{\rm in}}{h},
\end{equation}
one obtains from Eq. (\ref{eq:flux_disc})  $F_\nu \propto \nu^{1/3}$, known sometimes as the ``disc" spectrum, but such a $\nu^{1/3}$ feature is prominent
only if  $T({R_{\rm out}}) \ll  T_{\rm in}$ , i.e., if the disc is sufficiently large. 
For an AGN disc with $R_{\rm out}\sim 100 R_{\rm in}$ and $\dot m=0.01$ the
optical an UV spectrum will indeed have a 1/3 slope.

\subsection{Size of the disc}

Assuming that thin discs in AGNs emit locally as black-bodies allows one to determine the disc's size and comparing it with observation.
Defining the disc size as corresponding to the radius at which the disc temperature matches the wavelength $R_\lambda$:
\begin{equation}
\label{eq:rlambda}
kT(R_\lambda) = hc/\lambda,
\end{equation}
and using Eq. (\ref{eq:bbdisc}) in the form of
\begin{equation}
I_{\nu}=\frac{2 h_{p} c}{\lambda^{3}}\left[\exp \left(\frac{R}{R_{\lambda_{\text {rest}}}}\right)^{3 / 4}-1\right]^{-1},
\end{equation}
one obtains from Eqs.  (\ref{eq:Teff_value}) and (\ref{eq:Teff_in}) 
\begin{equation}
R_{\lambda}=\left[\frac{45 G \lambda^{4} M \dot{M}}{16 \pi^{6} h c^{2}}\right]^{1 / 3}=2.1 \times 10^{15}\left(\frac{\lambda}{\mu \mathrm{m}}\right)^{4 / 3}m_8^{2 / 3}\left(\frac{L}{\eta L_{E}}\right)^{1 / 3} \mathrm{cm}.
\end{equation}
$\lambda$ is the wavelength in the rest-frame of the AGN. Thus the prediction of the thin-disc model is that the size of the disc $R \sim M^{2/3}$.

Eq. (\ref{eq:rlambda}) assumes that emission at wavelength $\lambda$ originates solely at 
radius $R_\lambda$, while in reality this emission comes also from other radii.
Therefore a more appropriated size for comparison with observations would be a flux-weighted mean radius $\langle R_\lambda \rangle = X R_\lambda$.
$X \approx 2 - 3$, but if disc variability is taken into account this factor can be even $\sim 5$. $R_\lambda$ itself depends on the black-hole mass and the accretion rate
(through $\eta$) so that uncertainties in the observed values of this quantities impact the comparison of the model-size with observations. AGN disc sizes are measured
through reverberation mapping and microlensing. Observations are in agreement with the predicted $\sim M^{2/3}$ slope. There is no agreement about the actual size (see, e.g. \citealt{Jha0422} for a recent result). While
microlensing observations and some reverberation estimates claim values 2 -- 3 larger than those predicted by the model, according to other reverberation observations most
measured sizes of AGN discs agree with the thin-disc value. The reasons for the discrepancies (if real) is not clear, but it is possible that the real temperature profile is
shallower than the thin-disc $R^{-3/4}$ which should be the case when the disc is irradiated.

\subsection{Spectral lines from Keplerian accretion discs}
\label{sec:lines}
\begin{figure}
\center
\includegraphics[width=0.35\textwidth]{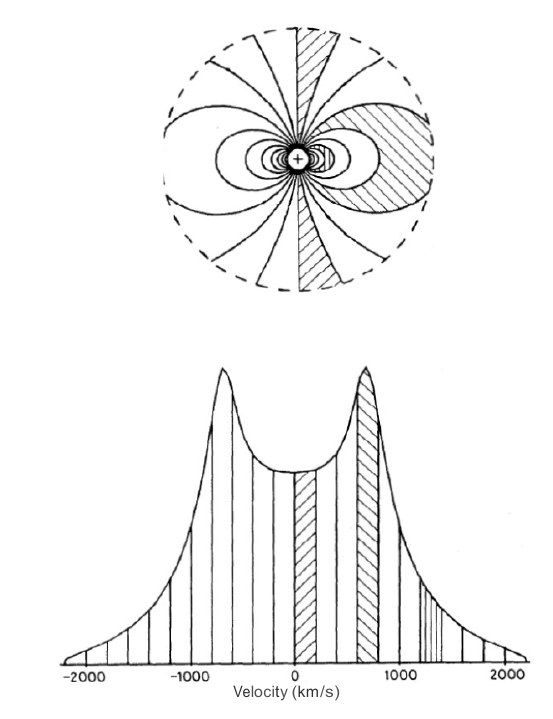}
\caption{The loci of constant radial velocity form a dipole field pattern on the surface of a Keplerian disc 
	(upper figure).
The velocity profile of emission lines from the
	disc (lower figure). Emission in the shaded velocity bins arises from the corresponding regions of the disc. This example corresponds to a stellar-mass accretor. [\textsl{Adapted from \cite{HM86}}]}
\label{fig:lineKepl}
\end{figure}
Observations of spectral lines from discs around black holes play an important role as tools allowing to investigate
the disc structure itself but their shapes encode the properties of the accretor, in this case its mass and spin. In practice
this means taking into account the relativistic effects (both special and general) on the propagation of light-rays which
might obfuscate the basic properties of line shapes emitted by an accretion disc. We will therefore begin with the Newtonian case.

Emission lines from a Keplerian accretion discs have a characteristic double-peak structure.
The reason is easy to understand. For a distant observer the frequency of the lines emitted by the disc are shifted
by the Doppler effect, corresponding to velocity
\begin{equation}
\label{eq:vd}
v_{\mathrm{D}}=v_{\mathrm{K}} \sin i \sin \theta,
\end{equation}
where $v_K= R\Omega_K$, $i$ is the disc inclination ($i=90^{\circ}$ for a disc seen edge-on) and $\theta$ is
the azimuth angle relative to the line-of-sight. The total disc emission flux is obtained by integrating the
Doppler-shifted intensities over the whole disc's surface:
\begin{equation}
\label{eq:fnu}
F_{\nu}=\frac{1}{D^{2}}\int {I_{\nu}} R\,{dR}d\theta.
\end{equation}
Here the line specific intensity is $I_\nu= j(R)\phi_\nu$, with 
$\int \phi_\nu\,d\nu=1$, $\phi_\nu$ being the emission profile of the line.
$j(R)$ is assumed to be isotropic. Each local line profile $I_\nu$ is shifted from the rest frequency
$\nu^{\prime}$ to the  local Doppler frequency
\begin{equation}
\label{eq:vdopp}
\nu_{\mathrm{D}}=\nu^{\prime}\left(1-\frac{v_{\mathrm{D}}}{c}\right).
\end{equation}
As seen on Figure \ref{fig:lineKepl} the loci of constant radial velocity form a ``dipole" pattern on the disc's surface,
The disc's emission-line profile, shown in the lower part of this figure is divided into velocity bins corresponding to the disc regions between consecutive dipole field lines.
The emission in each velocity bin arises from a different region of the disc surface. 
Assuming circular orbits, matter crossing the $\theta =0, \pi$ line moves perpendicularly to the line of sight, so 
$v_D=0$ and its emission correspond to the line center. 

The crescents that are complete near the accretor become truncated at the outer disc radius $R_D$ which produces 
cusps in the line profile at $v_D =\pm v_K\left(R_D\right)\,\sin i$. Within the assumptions
of a Keplerian potential, and an almost constant disk surface density, \textsl{double-peaked line profiles thus correspond to the
presence of an outer boundary to the emission-line region}. 

The wings of the lines are produced obviously by the fastest moving disc matter, i.e. close to the accretor.
However, in a Keplerian disc, the area of the corresponding crescent $R\Delta R \sim v^{-5}\Delta v$ which explains its rapid decline with increasing velocity.

\subsubsection{General-relativistic spectral line description}

Equations (\ref{eq:vd}), (\ref{eq:fnu}) and (\ref{eq:vdopp}) cannot describe correctly the spectral line close to a black hole
because in their derivation, higher-order terms in $v/c$ and gravitational light-bending have been neglected. The description of
such line profiles requires the use of the full formalism of General Relativity and usually requires numerical calculations. 
It what follows we will show how the problem can be simplified even in the case of a Kerr black hole (based on \citealt{HMP94}).
Exceptionally in this subsection we will use units in which c=G=1 in the description of the motion of the emitter.

The specific intensity [erg cm$^{-2}$ s$^{-1}$ Hz$^{-1}$ Sr$^{-1}$], is now written as
\begin{equation}
I_{\nu}=I_{\nu^{\prime}}(1+z)^{-3},
\end{equation}
where $\nu^{\prime}$ is the rest-frame frequency, $\nu$ the frequency 
measured at infinity and $1+z = \nu^{\prime}/\nu$ is the redshift factor.
One of the redshift factors in the denominator accounts for the time dilation, while the other two correspond to the angular diameter correction
to the solid angle. Therefore the observed spectral flux is
\begin{equation}
\label{eq:fnugr}
F_{\nu}=\underset{{\text {image }}} {\int} I_{\nu} d \Omega= \underset{{\text {image }}} {\int} I_{\nu^{\prime}}(1+z)^{-3} d \Omega,
\end{equation}
where $d\Omega$ is the solid angle and the integration is over the observed accretion disc image.

Therefore to calculate the observed line profiles one has to calculate the redshift factor $1+z$ which is given by the motion of the emitter in the
gravitational field of the attracting body. In the case of a Keplerian motion around a Kerr black hole, with orbital frequency
\begin{equation}
\label{eq:omegakgr}
\Omega_K=\frac{M^{1 / 2}}{R^{3 / 2}+a M^{3 / 2}},
\end{equation}
the result is
\begin{align}
\label{eq:redshiftgeneral}
& 1+z  = \\
& {\frac{1-L_z \Omega_K}{\sqrt{\left(1-2{M}/{R}-R^{2} \Omega_K^{2}\right)+4 a\Omega_K{M^2}/{R}  -\left(1+2 {M}/{R}\right) (aM)^{2} \Omega_K^{2}}}},
\end{align}
where $L_z$ is the component of the angular momentum of the emitter with respect to the $z$ (black-hole rotation) axis.
\begin{figure}
\centering
   \centering
    \includegraphics[width=0.7\textwidth]{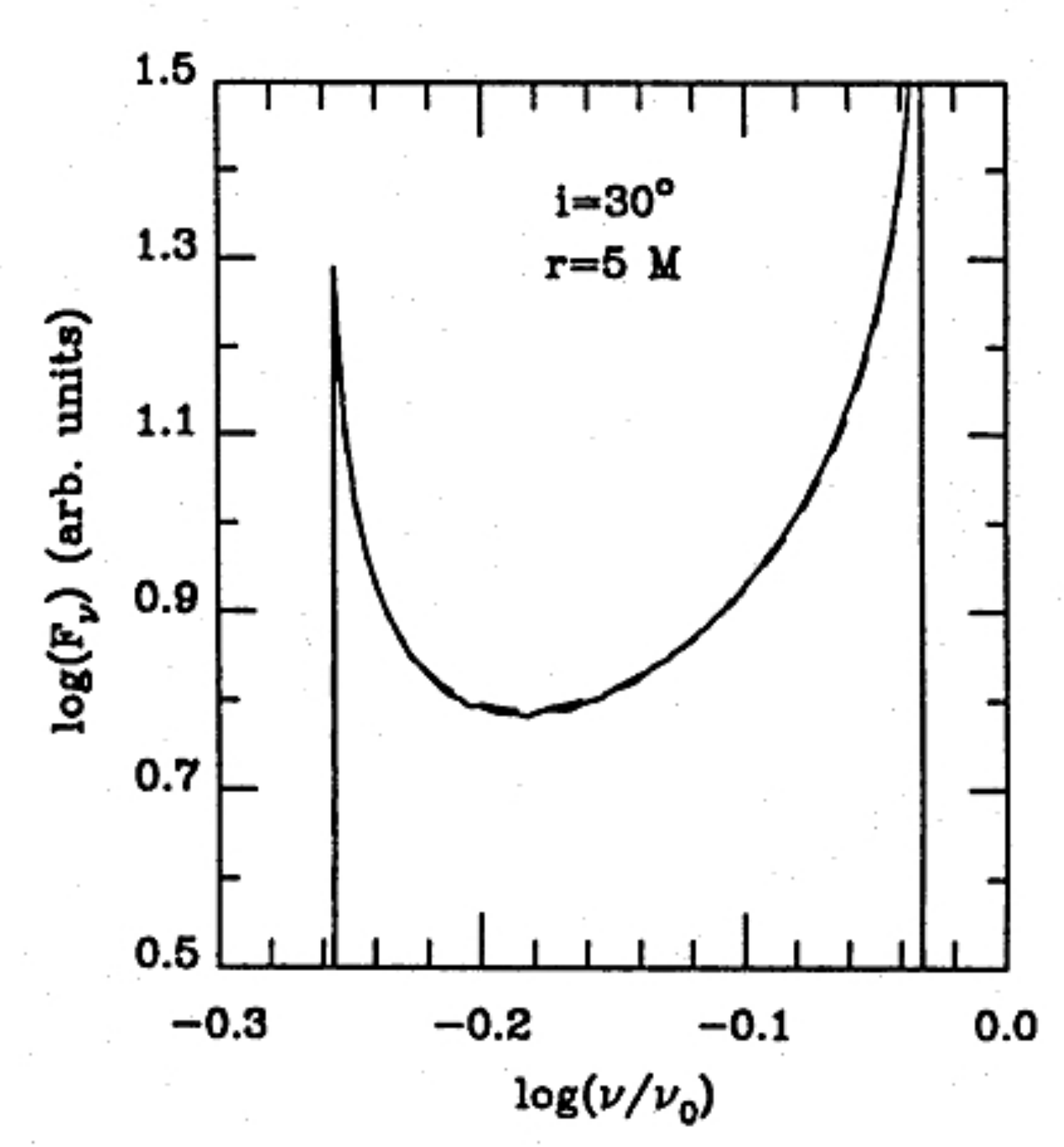}
	\caption{Exact (solid curve) and approximate (dashed line)  line profiles emitted  by a ring of matter at $r=2.5$ ($R=5GM/c^2$). The disc inclination is $30^{\circ}$. [\textsl{Adapted from \cite{HMP94}}]}
    \label{fig:Line1}
    \end{figure}
In general, Eq. (\ref{eq:fnugr}) with (\ref{eq:redshiftgeneral}) has to be solved numerically. However, for distances  $R > 10$ ($r > 5$) 
one can neglect the effects of black-hole rotation (they diminish as  $R^{-3}$; see Eq. \ref{eq:LTomega}) and Eqs. (\ref{eq:omegakgr}) and (\ref{eq:redshiftgeneral})
simplify to its Schwarzschild form
\begin{equation}
\Omega_K=\sqrt{\frac{M}{R^{3}}} \quad 1+z=\frac{1-\Phi \Omega_K}{\sqrt{1-3 M / R}}.
\end{equation}
Then, for low inclinations ($i \sim 0$) and large radii, one can easily integrate Eq. (\ref{eq:fnugr}), with $d\Omega \approx \cos i \,rdr\, d\varphi/4\pi D^2$.
Assuming a Dirac-delta intrinsic line profile $I_{\nu^{\prime}}= j(r) \delta(\nu^{\prime} - \nu_0)$ and defining
\begin{equation}
\alpha:=\sqrt{1-3 / \tilde{r}}, \quad \beta:=\sin i / \sqrt{\tilde{r}}, \quad \tilde{r} := R / M,
\end{equation}
one obtains
\begin{equation}
F_{\nu}=\frac{\cos i}{2 \pi D^{2} \nu_{0}} \int_{R_{\mathrm{in}}}^{R_{\mathrm{out}}} \frac{\alpha\left(\nu / \nu_{0}\right)^{3}  \phi(R) R\,dR}{\left(\beta^{2}\left(\nu / \nu_{0}\right)^{2}-\left(\alpha-\nu / \nu_{0}\right)^{2}\right)^{1 / 2}},
\end{equation}
which is an asymptotic formula for the spectrum observed at large distances from the black hole.
Close to the black hole, this formula is inappropriate even for low inclinations, since it neglects both
the effects of black-hole rotation and terms of the order of $\left( v/c\right)^3$. 
However, it happens to provide
an excellent approximation to the observed line profiles if $\alpha$ and $\beta$ are redefined as
\begin{equation}
\begin{aligned}[b]
\alpha &=\alpha_{0}\\
\beta &=\beta_{0}\left[1+\frac{1.25}{\tilde{r}}+\frac{2.5-3.4 \cos i}{\tilde{r}^{2}}+\frac{3 \sin i}{\tilde{r}^{3}}\right],
\end{aligned}
\end{equation}
for a Schwarzschild metric ($a=0$), and as
\begin{equation}
\begin{aligned}[b]
{\alpha=\alpha_{0}\left[1+\frac{5(\cos i-1)}{\tilde{r}^{5 / 2}}+\frac{6.5-8.5 \cos i}{\tilde{r}^{7 / 2}}\right]} \\ 
{\beta=\beta_{0}\left[1+\frac{1.25}{\tilde{r}}+\frac{3-3 \cos i}{\tilde{r}^{2}}-\frac{3.5 \sin i}{\tilde{r}^{3}}\right]},
\end{aligned}
\end{equation}
for a Kerr black hole with $a=0.99$. $\alpha_0=(1 - 3v^2)$, $\beta_0=v \sin i$ and $v=R\Omega_K$, with $\Omega_K$
given by Eq. (\ref{eq:omegakgr}).
 \begin{figure}
 \centering
    \includegraphics[width=0.7\textwidth]{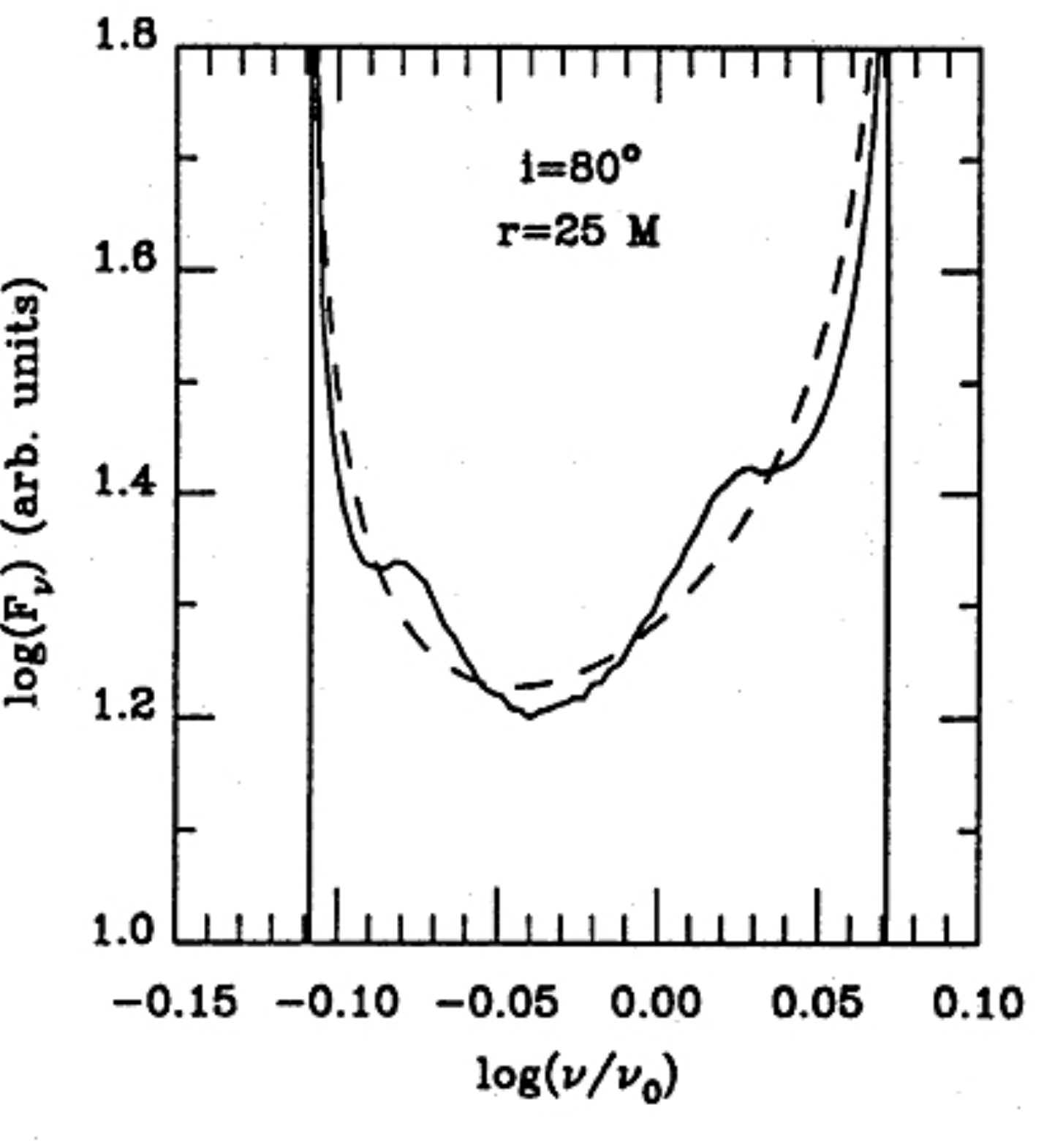}
    \caption{Same as in Fig. \ref{fig:Line1}, but for $r=12.5$ ($R=25GM/c^2$) and inclination $80^{\circ}$. For such high disc inclinations
	 the effects of light-bending (producing the two additional inner peaks) cannot be neglected even for relatively large radii. [\textsl{Adapted from \cite{HMP94}}]}
    \label{fig:Line2}
 \end{figure}
$\alpha$ and $\beta$ are related to the minimum and maximum frequencies of the observed line through
\begin{equation}
\begin{aligned}[b]
\alpha &=\frac{2 \nu_{\min } \nu_{\max }}{\left(\nu_{\min }+\nu_{\max }\right) / \nu_{0}} \\ 
\beta &=\frac{\nu_{\max }-\nu_{\min }}{\left(\nu_{\min }+\nu_{\max }\right) / \nu_{0}}.
\end{aligned}
\end{equation}
The fit to $\alpha$ and $\beta$ is accurate to better than $1\%$ for $r > 1.5$ in the Schwarzschild metric and 
$r > 2.5$ for a Kerr black hole, if $i < 80^{\circ}$.
Finally, the flux as given by Eq. (\ref{eq:fnugr}) has to be renormalised through factors that depend on the radius and inclination.
These factors are
\begin{equation}
\begin{aligned} 
C=1 &+\frac{1.25(\cos i-\sin i)+1.6 \tan i}{\tilde{r}^{1-0.19 \sin i}} \\ 
&+\frac{10-10 \cos i-3.5 \tan i}{\tilde{r}^{2}},
\end{aligned}
\end{equation}
for $a=0$
\begin{equation}
\begin{aligned} 
C=1 &+\frac{1.5-2 \sin i+1.2 \tan i}{\tilde{r}^{1-0.25 \sin i}}+\frac{7-7.6 \cos i+2 \sin i}{\tilde{r}^{2}} \\ 
&-\frac{3.5 \sin i}{\tilde{r}^{3}},
\end{aligned}
\end{equation}
for $a=0.99$. Multiplying by these factors gives the total line fluxes correct within less that $1\%$ for $i < 80^{\circ}$
when $r > 1.5$ in the Schwarzschild metric and  and $r > 2.5$ for a Kerr black hole. $\tan i$ in these formulae represents the light
bending effect.

Figures \ref{fig:Line1} and \ref{fig:Line2} show two examples of  ``exact" (numerical) and approximated fitting of
the profile of a line emitted by a ring around a $a=0.99$ black hole. For a low inclination ($i = 30^{\circ}$) even
close to the black hole ($r=2.5$) the numerical and approximate profiles are practically indistinguishable. However,
even at relatively large distance from the hole ($r=12.5$) but high inclination ($i = 80^{\circ}$) the imperfections
of the approximate version of the flux are clearly visible.

The double-peaked line shape is, as in the newtonian case, due to the finite extent of the emission region, in this case a ring.
The often used assumption $j(R) \sim R^{-b}$ will not produce double-peak lines if the outer emission region is not spatially
limited.

\subsection{Radiative structure}
\label{subsect:rad}

Here we will show an example of the solution for the vertical thin--disc structure
which exhibits properties impossible to identify when the structure is vertically
averaged. We will also consider here irradiated discs -- such as AGN discs.

We write the energy conservation as :
\begin{equation}
\frac{dF}{ dz} = q^+(R,z),
\label{energy1}
\end{equation}
where $F$ is the vertical (in the $z$ direction) radiative flux 
and $q^+(R,z)$ is the viscous heating rate per unit volume.
Eq. (\ref{energy1})
states that an accretion disc is not in radiative equilibrium ($dF/dz\neq 0$), contrary to a
stellar atmosphere. For this equation to be solved, the function 
$q^+(R,z)$ must be known. As explained and discussed in Sect. \ref{subsec:vstruct.1} the viscous dissipation is often written as 
\begin{equation}
q^+(R,z)= \frac{3}{2} \alpha \Omega_{\rm K} P(z).
\label{voldiss}
\end{equation}
Viscous heating of this form has important implications for the structure of
optically thin layers of accretion discs and may lead to the creation of
coronae and winds. In reality it is an \textsl{ad hoc} formula inspired by
Eq. (\ref{eq:qplusalphaP}). We don't know yet how to describe the viscous
heating stratification in a real \textsl{geometrically thin} accretion disc and Eq. (\ref{voldiss}) just
\textsl{assumes} that it is proportional to pressure. It is simple and convenient
but it is not necessarily true.

When integrated over $z$, the rhs of Eq. (\ref{energy1}) using Eq. (\ref{voldiss})
is equal to viscous dissipation per unit surface:
\begin{equation}
F^+=\frac{3} {2} \alpha \Omega_{\rm K} \int_0^{+\infty} P dz ,
\label{fvis}
\end{equation}
where $F^+=(1/2)Q^+$ because of the integration from $0$ to $+\infty$ while
$Q^+$ contains $\Sigma$ which is integrated from $-\infty$ to $+\infty$ (Eq. \ref{eq:Sigma}).

One can rewrite Eq. (\ref{energy1}) as
\begin{equation}
\frac{dF}{ d\tau} = - f(\tau )\frac{F_{\rm vis} }{ \tau_{\rm tot}},
\label{energy2}
\end{equation}
where we introduced a new variable, the optical depth $d\tau=-\kappa_{\rm R}
\rho dz$, $\kappa_{\rm R}$ being the Rosseland mean opacity and $\tau_{\rm tot}
= \int_0^{+\infty} \kappa_{\rm R} \rho dz$ is the total optical depth.
$f(\tau)$ is of order unity.

At the disc midplane, by symmetry, the flux must vanish: $F(\tau_{\rm tot})=0$,
whereas at the surface, ($\tau=0$)
\begin{equation}
F(0) \equiv \sigma T^4_{\rm eff}= F^+.
\label{fsurface}
\end{equation}
Equation (\ref{fsurface}) states that the total flux at the surface is equal
to the energy dissipated by viscosity (per unit time and unit surface). The
solution of Eq. (\ref{energy2}) is thus
\begin{equation}
F(\tau) = F^+ \left(1 - \frac{\int_0^\tau f(\tau) d\tau}{\tau_{\rm tot}}\right),
\label{flux0}
\end{equation}
where $\int_0^{\tau_{\rm tot}} f(\tau) d\tau = \tau_{\rm tot}$. Since
$f \approx 1$, we have
\begin{equation}
F(\tau) \approx F^+ \left(1 - \frac{\tau}{\tau_{\rm tot}}\right).
\label{fluxd}
\end{equation}

To obtain the temperature stratification one has to solve the transfer
equation. Here we use the diffusion approximation
\begin{equation}
F(\tau) = \frac{4}{3} \frac{\sigma dT^4}{d\tau} ,
\label{diff}
\end{equation}
appropriate for the optically thick discs we are dealing with. The
integration of Eq. (\ref{diff}) is straightforward and gives :
\begin{equation}
T^4(\tau) - T^4(0) = \frac{3}{4} \tau \left(1 - \frac{\tau}{2\tau_{\rm tot}}
            \right) T^4_{\rm eff}.
\label{t1}
\end{equation}

The upper (surface) boundary condition is:
\begin{equation}
T^4(0) = \frac{1}{2} T^4_{\rm eff} + T^4_{\rm irr},
\label{bcond2}
\end{equation}
where $T^4_{\rm irr}$ is the irradiation temperature, which depends on $r$,
the albedo, the height at which the energy is deposited and on the shape of
the disc. In Eq. (\ref{bcond2}) $T(0)$ corresponds to
the {\sl emergent} flux and, as mentioned above, $T_{\rm eff}$ corresponds to
the {\sl total} flux ($\sigma T^4_{\rm eff}=Q^+$) which explains the factor 1/2
in Eq (\ref{bcond2}). The temperature stratification
is thus :
\begin{equation}
T^4(\tau) = \frac{3}{4}T^4_{\rm eff}
             \left[\tau \left(1 - \frac{\tau}{2\tau_{\rm tot}}\right)
        + \frac{2}{3}\right] + T^4_{\rm irr}.
\label{t2}
\end{equation}
For $\tau_{\rm tot} \gg 1$ the first
term on the rhs has the form familiar from the stellar atmosphere models in the
Eddington approximation.

\textsl{In this case at $\tau=2/3$ one has $T(2/3) = T_{\rm eff}$.}

Also for $\tau_{\rm tot} \gg 1$, the temperature
at the disc midplane is
\begin{equation}
T^4_{\rm c} \equiv T^4(\tau_{\rm tot}) =
         \frac{3}{8} \tau_{\rm tot} T_{\rm eff}^4 + T^4_{\rm irr}.
\label{diff2}
\end{equation}
It is clear therefore that for the disc inner structure to be dominated by
irradiation and the disc to be isothermal one must have
\begin{equation}
\frac{F_{\rm irr}}{\tau_{\rm tot}} \equiv \dfrac{}{}{\sigma T^4_{\rm irr}}{\tau_{\rm tot}} \gg F^+
\label{c1}
\end{equation}
and not just $F_{\rm irr} \gg F^+$ as is sometimes assumed. The difference between the two criteria is important in X-ray binary and AGN discs since, for parameters of interest, $\tau_{\rm tot} \gtrsim 10^2 - 10^3$ (see Sect. \ref{subsec:scurve}).

\subsection{Shakura-Sunyaev thin-disc solution}
\label{subsec:SS}

In their seminal and famous 1973 paper, Shakura \& Sunyaev found power-law stationary solutions of the simplified version of the thin--disc equations presented in Sects. \ref{subsec:vstruct.1}, \ref{subsec:radial1} and \ref{subsec:stationary}. The 8 equations for the 8 unknowns $T_c$, $\rho$, $P$, $\Sigma$, $H$, $\nu$, $\tau$ and $c_s$ can be written as
\begin{equation}
\label{}
  \Sigma=2H\rho \ \ \ \  \ \ \  \ \ \ \ \ \ \ \ \ \ \ \ \ \ \ \ \ \ \ \ \ \ \ \ \ \ \ \ \ \ \ \ \ \ \ \ \ \ \ \ \ \ \ \ \ \ \ \ \ \ \ \ \ \ \ \ \ \rm (\i) \nonumber
\end{equation}
\begin{equation}
\label{}
 H=\frac{c_s R^{3/2}}{(GM)^{1/2}} \\ \ \ \ \ \ \ \ \ \ \ \ \ \ \ \ \ \ \ \ \ \ \ \ \ \ \ \ \ \ \ \ \ \ \ \ \ \ \ \ \ \ \ \ \ \ \ \ \ \ \ \ \ \ \ \ \ \ \ \rm (\i\i) \nonumber
\end{equation}
\begin{equation}
\label{}
c_s=\sqrt{\frac{P}{\rho}}\\ \ \ \ \ \ \ \ \ \ \ \ \ \ \ \ \ \ \ \ \ \ \ \ \ \ \ \ \ \ \ \ \ \ \ \ \ \ \ \ \ \ \ \ \ \ \ \ \ \ \ \ \ \ \ \ \ \ \ \ \ \ \ \ \ \ \rm (\i\i\i) \nonumber
\end{equation}
\begin{equation}
\label{}
P=\frac{{\cal R}\rho T}{\mu} + \frac{4\sigma}{3c}T^4 \\ \ \ \ \ \ \ \ \ \ \ \ \ \ \ \ \ \ \ \ \ \ \ \ \ \ \ \ \ \ \ \ \ \ \ \ \ \ \ \ \ \ \ \ \ \ \ \ \ \ \ \ \ \ \ \rm(\i v) \nonumber
\end{equation}
\begin{equation}
\label{}
\tau(\rho, \Sigma, T_c)=\kappa_R(\rho,T_c)\Sigma \\ \ \ \ \ \ \ \ \ \ \ \ \ \ \ \ \ \ \ \ \ \ \ \ \ \ \ \ \ \ \ \ \ \ \ \ \ \ \ \ \ \ \ \ \ \ \ \rm(v) \nonumber
\end{equation}
\begin{equation}
\label{}
\nu(\rho,\Sigma,T_c,\alpha)=\frac{2}{3}\alpha c_s H   \\ \ \ \ \ \ \ \ \ \ \ \ \ \ \ \ \ \ \ \ \ \ \ \ \ \ \ \ \ \ \ \ \ \ \ \ \ \ \ \ \ \ \ \ \ \ \ \ \ \rm(v\i) \nonumber
\end{equation}
\begin{equation}
\label{}
\nu \Sigma =\frac{\dot M }{3\pi}\left[1 -\left(\frac{R_{\mathrm in}}{R}\right)^{1/2} \right] \\ \ \ \ \ \ \ \ \ \ \ \ \ \ \ \ \ \ \ \ \ \ \ \ \ \ \ \ \ \ \ \ \ \ \ \ \ \ \ \ \ \ \ \ \ \rm(v\i\i) \nonumber
\end{equation}
\begin{equation}
\label{}
\frac{8}{3}\frac{\sigma T_c^4}{\tau}=\frac{3}{8\pi}\frac{GM\dot M}{R^3}\left[1 -\left(\frac{R_{\mathrm in}}{R}\right)^{1/2} \right] . \\ \ \ \ \ \ \ \ \ \ \ \ \ \ \ \ \ \ \ \ \ \ \ \ \ \ \ \ \ \ \ \rm(v\i\i\i) \nonumber
\end{equation}
Equations (\i) and (\i\i) correspond to vertical structure equations (\ref{eq:mass_cons}) and (\ref{eq:mec_approx}), Eq. (v\i\i) is the radial Eq. (\ref{eq:amintK}), while Eq. (v\i\i\i) connects vertical to radial equations. Eq. (\i\i\i) defines the sound speed, Eq. (\i v) is the equation of state and (v\i) contains the information about opacities. The viscosity $\alpha$ parametrization introduced in \cite{SS73} provides the closure of the 8 disc equations. Therefore they can be solved for a given set of $\alpha$, $M$, $R$ and $\dot M$.

Power-law solutions of these equations exist in physical regimes where the opacity can be represented in the  Kramers form $\kappa=\kappa_0\rho^{n}T^{m}$ and one of the two pressures, gas or radiation, dominates over the other. There are three regimes to be considered: 
\begin{center}
$\left. a. \right)$ $P_r \gg P_g$ and $\kappa_{\rm es}\gg \kappa_{\rm ff}$\\
$\left. b. \right)$ $P_g \gg P_r$ and $\kappa_{\rm es}\gg \kappa_{\rm ff}$ \\
$\left. c. \right)$ $P_g \gg P_r$ and $\kappa_{\rm ff}\gg \kappa_{\rm es}$.\\
\end{center}
Regimes $\left. a. \right)$ and $\left. b. \right)$ in which opacity is dominated by electron scattering will be discussed in Sect. \ref{sect:advection}.
Here we will present the solutions of regime $\left. c. \right)$, i.e. we will assume that
\begin{equation}
\label{eq:kramers}
P_r=0 \ \ \ \ {\mathrm{and}} \ \ \ \ \kappa_R=\kappa_{\rm ff}=5\times 10^{24}\rho T_c^{-7/2}\,\rm cm^2g^{-1}.
\end{equation}

The solution for the surface density $\Sigma$, central temperature $T_c$ and the disc relative height (aspect ratio) are respectively
\begin{equation}
\label{eq:SigmaSS}
\Sigma=1.6\times 10^7\,\alpha^{-4/5}m_8^{+1/5}r^{-3/4}{\dot m}^{7/10}f^{7/10}\, \rm g\,cm^{-2},
\end{equation}
\begin{equation}
\label{eq:TcSS}
T_c=8.4\times 10^6\,\alpha^{-1/5}m_8^{-1/5}r^{-3/4}{\dot m}^{3/10}f^{3/10}\,\rm K,
\end{equation}
\begin{equation}
\label{eq:HoverRSS}
\frac{H}{R}= 1.6\,\times 10^{-3}\alpha^{-1/10}m_8^{-1/10}r^{1/8}{\dot m}^{3/20}f^{3/20},
\end{equation}
where $m_8=M/{\rm 10^8 M_{\odot}}$, $r=R /R_S$, and
$f~=~1~ -~ ({R_{\rm in}}/{R})^{1/2}$.

%Since
%\begin{equation}
%\frac{P_g}{P_r}= 1.6 \times 10^{-2} \alpha^{-1/10}m_8^{1/2}r^{3/8}\dot m^{-7/20}f^{-7/20},
%\label{eq:pgpr}
%\end{equation}
%the regime c.) solution exists only for radii
%\begin{equation}
%r > 6\times 10^4 \alpha^{4/15}m_8^{4/3}\dot m^{14/15}f^{14/15}.
%\label{eq: rPr}
%\end{equation}

Since
\begin{equation}
\frac{P_g}{P_r}= 0.32\, \alpha^{-1/10}m_8^{-1/10}r^{3/8}\dot m^{-7/20}f^{-7/20},
\label{eq:pgpr}
\end{equation}
the regime c.) solution exists only for radii
\begin{equation}
r \gtrsim 21 \alpha^{4/15}m_8^{4/15}\dot m^{14/15}.
\label{eq: rPr}
\end{equation}

For this solution the viscous time is
\begin{equation}
t_{\rm vis}\approx 26\, \alpha^{-6/5} m_8^{6/5}\dot m^{-3/8}r^{5/4} \rm yr,
\label{eq:viscsol}
\end{equation}
so for the viscous time to be less than the age of the Universe (13.8 Gyr), for an Eddington accretion rate onto a $10^8\Msun$ black-hole, the outer disc radius must be
$r_{\rm out}< 10^7 \alpha^{24/25}$, i.e. less than $\approx 100$\,pc, or less than $1\,\rm pc$ if $\alpha$ in AGNs is $\sim 10^{-2}$.

In this solution it was assumed that the opacity is given by the formula
\begin{equation}
\kappa_{\rm R}=5\times 10^{-4}\dot m^{-1/2} r^{3/4} \rm cm^2g^{-1},
\label{eq:kappass}
\end{equation}
so that $\kappa_{\rm R} > \kappa_{\rm es}=0.4\,\rm cm^2g^{-1}$, for
\begin{equation}
r_{\rm Res} > 7400 \dot m^{2/3}.
\label{eq:rkappases}
\end{equation}
Notice that $r_{\rm Res}$ is independent on the accretor's mass.

It is characteristic of the Shakura-Sunyaev solution in this regime that the three $\Sigma$, $T_c$ and $T_{\rm eff}$ radial profiles vary as $ R^{-3/4}$. 
(This implies that the optical depth $\tau$ is constant with radius $--$ see Eq. v\i\i\i.) 
For high accretion rates and small radii, the
assumption of opacity dominated by free-free and bound-free absorption will brake down, and the solution will cease to be valid. 
We will first consider the other disc end: large radii. 

There, the temperature will finally drop ($T_c \sim T^{-3/4}$) below $10^4$ K, the disc plasma will recombine leading to a drastic change in opacities which triggers a thermal instability.

\section{Disc instabilities}
\label{sec:DI}

In this section we will present and discuss the disc thermal and the (related) viscous instabilities.
In AGN discs two instabilities are of interest: one is related to the hydrogen ionisation/recombination,
the other occurs in discs where pressure is dominated by radiation. The instability is thermal when it grows
on a thermal timescale, in a geometrically thin disc it means that the timescale is $ \sim (\alpha\,\Omega_K)^{-1}$
(see Eq. \ref{eq:thermal_time}).

 \subsection{The thermal instability}
 \label{subsec:thermal_instability}

A disc is thermally stable if the radiative cooling varies with
temperature faster than the viscous heating. In other words a disc is thermally stable if
\begin{equation}
\frac{d\ln \sigma T_{\rm eff}^4}{ d\ln T_{\rm c}} > \frac{d\ln Q^+}{d\ln T_{\rm c}}.
\label{stab}
\end{equation}
Using Eq. (\ref{diff2}) one obtains
\begin{equation}
\frac{d\ln T_{\rm eff}^4}{ d\ln T_{\rm c}} =
4\left[1 - \left(\frac{T_{\rm irr}}{T_{\rm c}}\right)^{4}\right]^{-1} -
    \frac{d\ln \kappa}{d\ln T_{\rm c}}.
\label{eq:cool}
\end{equation}

One can see from these equations that disc stability is determined by the temperature dependence of
the opacities and that irradiation stabilises accretion discs.

\subsubsection{Thermal instability due to hydrogen ionisation/recombination}

In a gas--pressure dominated disc $Q^+ \sim \rho T\,H \sim \Sigma T \sim T_{\rm c}$ .
At high temperatures
${d\ln \kappa/d\ln T_{\rm c}}\approx - 4$ (see Eq. \ref{eq:kramers}).
A thermal instability arises due to a rapid change of opacities with temperature when hydrogen begins to recombine ($T_c \sim 10^4$K) and
this opacity dependence on temperature is no longer valid.  In the instability region, the
temperature exponent becomes large and \textsl{positive}: ${d\ln \kappa/ d\ln T_{\rm
c}} \approx 7 - 10$, and in the end cooling is decreasing with temperature.

Thus this thermal instability should be present in accretion discs with midplane temperature 
$T_c$ equal few times $10^4$K and below.
It has been established that it is at the origin of outbursts observed in discs around
stellar-mass black-holes, neutron stars and white dwarfs. Systems containing the first two
classes of objects are known as Soft X-ray transients (SXTs, where ``soft" relates
to their X-ray spectrum), while those containing white-dwarfs are called dwarf-novae
(despite the name that could suggest otherwise, nova and supernova outbursts
have nothing to do with accretion disc outbursts). This range of temperatures is also
present in discs around AGNs and we will present its consequences in Sect. \ref{sec:outbagn}.

\subsubsection{Thermal instability of radiation--pressure dominated discs}

Radiation--pressure dominated ($P=P_{\rm rad}\propto T^4$) accretion discs are thermally unstable when the
opacity is due to electron scattering on electrons. Indeed, in this case
\begin{equation}
\frac{d\ln T_{\rm eff}^4}{ d\ln T_{\rm c}} = 4
\label{eq:cooles}
\end{equation}
because $\kappa_R=\kappa_{\rm es}=const.$, while
in a radiation pressure dominated disc 
\begin{equation}
\label{eq:qradp1}
Q^+ \sim HT^4\sim T^8/\Sigma
\end{equation}
so that
\begin{equation}
\frac{d\ln Q^+}{d\ln T_{\rm c}} = 8 \, > \, \frac{d\ln T_{\rm eff}^4}{ d\ln T_{\rm c}}
\label{eq:heates}
\end{equation}
and the disc is thermally unstable. This solution is represented by the middle branch with negative slope on the S-curve in Fig. \ref{fig:adaf1} (see also Eq. \ref{eq:radpdom}) .

The presence of this instability in the model is one of the unsolved problems of the accretion disc theory because it contradicts observations which do not
show any unstable behaviour in the range of luminosities where discs should be in the radiative pressure and electron-scattering opacity domination regime.
This suggests that the disc description in this regime is incomplete.

For example, the radiation-pressure instability can be quenched if the disc vertical support is provided not by radiation but by a magnetic field. This field has to be brought into the disc
from the surrounding medium, because the MRI dynamo is unable to produce a field with the required strength.

The heating rate can be rewritten as
\begin{equation}
Q^{+}=  \alpha P_{\mathrm{rad}} H  \frac{\mathrm{d} \Omega}{\mathrm{d} r} = \alpha P_{\mathrm{rad}}^{2} \frac{\Sigma}{\Omega_{\mathrm{K}}^{2}} \frac{\mathrm{d} \Omega}{\mathrm{d} r},
\end{equation}
and from
\begin{equation}
\left.\frac{\mathrm{d} \log Q^{+}}{\mathrm{d} \log P_{\mathrm{rad}}}\right|_{\Sigma}=2;  \ \ \ \ \ \ \left.\frac{\mathrm{d} \log Q^{-}}{\mathrm{d} \log P_{\mathrm{rad}}}\right|_{\Sigma}=1,
\end{equation}
one deduces, as before, that a radiation-pressure dominated disc with electron-scattering opacity is 
thermally unstable.

However, with a magnetic-field pressure $P_{\mathrm{mag}}$, introducing 
\begin{equation}
\beta^{\prime}=\frac{P_{\mathrm{mag}}}{P_{\mathrm{rad}} + P_{\mathrm{mag}}}
\end{equation}
one can write the heating rate as
\begin{equation}
Q^{+}=\frac{\alpha P_{\mathrm{rad}}^{2}}{\left(1-\beta^{\prime}\right)^{2}} \frac{\Sigma}{\Omega_{\mathrm{K}}^{2}} \frac{\mathrm{d} \Omega}{\mathrm{d} r},
\end{equation}
from which one obtains

\begin{equation}
\left.\frac{\mathrm{d} \log Q^{+}}{\mathrm{d} \log P_{\mathrm{rad}}}\right|_{\Sigma, P_{\mathrm{mag}}}=2\left(1-\beta^{\prime}\right).
\end{equation}
Radiative cooling is independent of the magnetic pressure, so for $\beta^{\prime} > 0.5$ the heating
rate no longer increases more rapidly with the central radiation
pressure, than the cooling rate and the disc is thermally stable. It is not clear, however, how and 
from where a stabilising magnetic field will find itself in the right place and the right moment
to quench the disc's thermal instability (see a related discussion in Sect. \ref{sec:disc-jets}).

\subsection{Thermal equilibria: the \bS-curve}
\label{subsec:scurve}

We will consider thermal equilibria of an accretion disc in which
heating is due only to local turbulence, neglecting the
effects of irradiation, because in the case of AGN discs 
(contrary to discs in X-ray binaries) irradiation does not make much difference to
the instability's consequences. We put therefore $T_{\rm
irr}=\widetilde Q=0$. 
\begin{figure} [!]
\center
\includegraphics[angle=270,width=0.95\textwidth]{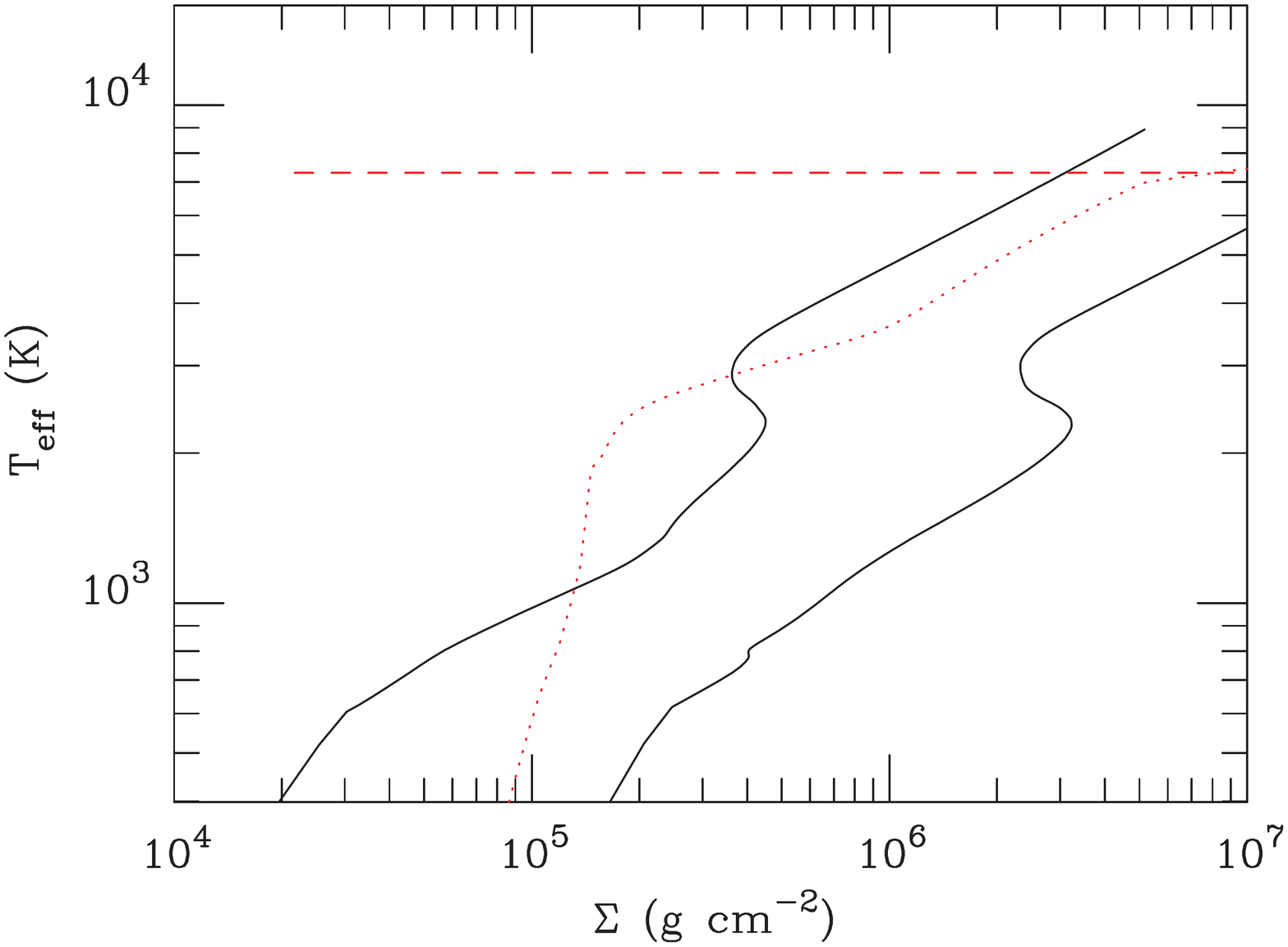}
  \caption{\bS-curves in the $\Sigma - T_{\rm eff}$ plane, for $M = 10^8 \msun$ and  $R= 2 \times 10^{16}$ cm. 
\bS-curves for two values of the viscosity parameter are shown: $\alpha = 0.1$ (left) and $\alpha=0.01$ (right). 
The dotted curve corresponds to $Q_T=1$ (see Eq. \ref{eq:toomre}), and the dashed curve to $H/R = 0.1$.
The model applies only to regions below the dashed curved and above the dotted one, i.e., to a geometrically thin ($H/R\ll 1$), non-self-gravitating  ($Q_T < 1$) accretion discs. In this case the thin disc approximation
applies to $r \lesssim 600$. [{\sl From \citet{HVL09}}].}
\label{fig:scurve}
\end{figure}

The thermal equilibrium in the disc is defined by the equation $Q^-=Q^+$
(see Eq. \ref{eq:heat}), i.e. by
\begin{equation}
\sigma T_{\rm eff}^4=\frac{9}{8} \nu \Sigma \Omega_{\rm K}^2
\label{termeq}
\end{equation}
(Eq. \ref{eq:vischeat2}).
In general, $\nu$ is a function of density and temperature and in the following we will use
the standard $\alpha$--prescription Eq.  (\ref{eq:nualphaP}). The energy
transfer equation  provides a relation between the
effective and the disc midplane temperatures so that thermal equilibria
can be represented as a $T_{\rm eff}\left(\Sigma\right)$~ --~relation 
(or equivalently a $\dot M(\Sigma)$--relation). In the relevant range of temperatures ($10^3 \lesssim T_{\rm eff}\lesssim 10^5$)
this relation forms an {\bS} on the
($\Sigma, T_{\rm eff}$) plane as in Fig. \ref{fig:scurve}. The upper, hot branch corresponds to the Shakura-Sunyaev solution
presented in Section \ref{subsec:SS}. The two other branches correspond to solutions for cold (partially ionised) discs. Near the upper bend of the {\bS} 
and below it, convection can play an important role in the energy transfer.

Each point on the ($\Sigma, T_{\rm eff}$) \bS-curve represents an
accretion disc's thermal equilibrium at a given radius, i.e. a thermal equilibrium of a ring at radius $R$. In other words
each point of the \bS-curve is a solution of the $Q^+=Q^-$ equation. 
Points not on the \bS-curve correspond to solutions of Eq. (\ref{eq:heat})  \textsl{out
of thermal equilibrium}: on the left of the equilibrium curve, cooling dominates over heating, $Q^+ < Q^-$; on the right 
heating over cooling $Q^+ > Q^-$.  Therefore a positive slope
of the $T_{\rm eff}(\Sigma)$ curve corresponds to \textsl{stable solutions}, since a small increase of temperature of an equilibrium state (an upward perturbation) on the upper branch, say, 
will bring the ring to a state where $Q^+ < Q^-$ so it will cool down getting back to equilibrium. In a similar way
a downward perturbation will provoke increased heating bringing back the system to equilibrium. 

The opposite is happening along the \bS-curve's segment with a negative slope as both a temperature increase and
decrease lead to a runaway. The middle branch of the \bS-curve corresponds therefore to \textsl{thermally unstable}
equilibria.

A stable disc equilibrium can be represented only by a point on the lower, cold or
the upper, hot branch of the \bS-curve. This means that the surface
density in a stable cold state must be \textsl{lower} than the maximal value on the
cold branch: $\Sigma_{\rm max}$, whereas the surface density in the hot
stable state must be \textsl{larger} than the minimum value on this branch: $\Sigma_{\rm
min}$. Both these critical densities are functions of the viscosity
parameter $\alpha$, the mass of the accreting object, the distance from the
center and depend on the disc's chemical composition. In the case of solar composition the critical surface densities are
\begin{equation}
\label{eq:Sigmacrit}
\Sigma_{\min }=2.90 \times 10^{3} \alpha^{-0.74}{R}_{15}^{1.04} m_{8}^{-0.35} \mathrm{g} \mathrm{cm}^{-2}
\end{equation}
\begin{equation}
\Sigma_{\max }=3.85 \times 10^{3} \alpha^{-0.82} {R}_{15}^{0.99} m_{8}^{-0.33} \mathrm{g} \mathrm{cm}^{-2}
\end{equation}
and the corresponing effective temperatures
\begin{equation}
\label{eq:Tcrit}
T_{\mathrm{eff}}\left(\Sigma_{\min }\right)=4300 R_{15}^{-0.12} \mathrm{K}
\end{equation}
\begin{equation}
\label{eq:Tcritplus}
T_{\mathrm{eff }}\left(\Sigma_{\max }\right)=3300 R_{15}^{-0.12} \mathrm{K},
\end{equation}
where $R=:R_{15}10^{15}\rm cm$.
Using the equation (see Eq. \ref{eq:teff})
\begin{equation}
\dot M= \sigma T_{\rm eff}^4\frac{8\pi R^3}{3GM},
\end{equation}
one obtains for the critical accretion rates
\begin{align}
 \dot M^+_{\rm crit}  :=    \dot M({\Sigma_{\rm min}})   = 1.22 \times 10^{22} R_{15}^{2.52}         \rm g\,s^{-1}   \label{eq:mdotcritplus}\\
 \dot M^-_{\rm crit}  :=    \dot M(\Sigma_{\rm max})   = 4.22 \times 10^{21} R_{15}^{2.52}         \rm g\,s^{-1} .  \label{eq:mdotcritminus}
 \end{align}

A stationary accretion disc in which there is a ring with effective temperature contained between the critical values of
Eq. (\ref{eq:Tcrit}) and (\ref{eq:Tcritplus}) cannot be stable. Since in a stationary disc, the effective temperature and the surface density both decrease with radius, the
stability of a disc depends on the accretion rate and the disc size. For a given
accretion rate a stable disc cannot have an outer radius larger than the value corresponding to Eq. (\ref{eq:mdotcritplus}).

A disc is hot and stable if the rate  at which mass is brought to its outer edge ($R\sim R_D$)  is larger than the critical accretion rate
at this radius $\dot{M}_{\rm crit}^{+}(R_D)$. On the other hand a disc is cold and stable if the mass--arrival rate to the disc is lower than the
critical accretion rate at its inner  radius $\dot{M}_{\rm crit}^{-}(R_{\rm in})$.

Taking $R_{\rm in} = 3 R_S$ and $R_D = 2\times 10^{16}\rm cm$, for $10^8\,\msun$ black hole, an accretion disc will be unstable if the mass--arrival rate is contained in the range
\begin{align}
9.6\times 10^{20}\, \rm g\,s^{-1} < \dot M < 2.3 \times 10^{25}\, \rm g\,s^{-1}\, \rm or, \label{eq:stab8}\\
6.8 \times 10^{-7}\, \msun\, \rm yr^{-1} < \dot M < 1.7 \times 10^{-2}\, \msun\, \rm yr ^{-1}
\end{align}
or
\begin{equation}
\label{eq:stab8edd}
6.0 \times 10^{-6} < \dot m < 0.6.
\end{equation}
This means that a typical accretion disc in an AGN \textsl{is thermally unstable}.

\subsection{Variability of unstable AGN discs}
\label{sec:outbagn}

The thermal instability at a certain disc's radius produces a steep temperature gradient. This in turn creates a heating front which
propagates radially bringing the cold disc regions to a hot state in which the heated up
disc's rings are on the upper branch of the \bS-curve. The rise of temperature increases the viscosity (Eq. \ref{eq:nualphaOmega}).
In the discs around stellar-mass objects, such as white dwarfs, neutron stars and black holes, this process redistributes the mass of the 
disc, creating a quasi-stationary disc configuration with an accretion rate $\dot M \approx \mathrm{const.} \approx \mathrm{few} \times \dot M^+_{\mathrm{crit}}(R_D)$,
the hot critical accretion rate at the outer disc radius. This corresponds to the outburst maximum, from which the accretion rate and luminosity decay
because, by construction, in an unstable disc the mass-feeding rate is less than the hot critical accretion rate $\dot M^+_{\rm crit}(R_D)$ (see Eq. \ref{eq:stab8}).
When the accretion rate drops below this value, a cooling-front forms in the outer disc regions and propagates towards the center, bringing the disc behind it to a cold state, thus switching-off the outburst.
During the following quiescence the disc, emptied during the outburst, fills-up again until it reaches somewhere the critical temperature, leading to the next outburst in the cycle.
A typical dwarf-nova outburst or an X-ray binary transient event is characterised by a fast rise and slow decay which correspond to a viscous decay of a disc with a shrinking outer radius.
\label{sec:variability}
\begin{figure}[!]
  \centering
  \includegraphics[width=0.7\textwidth]{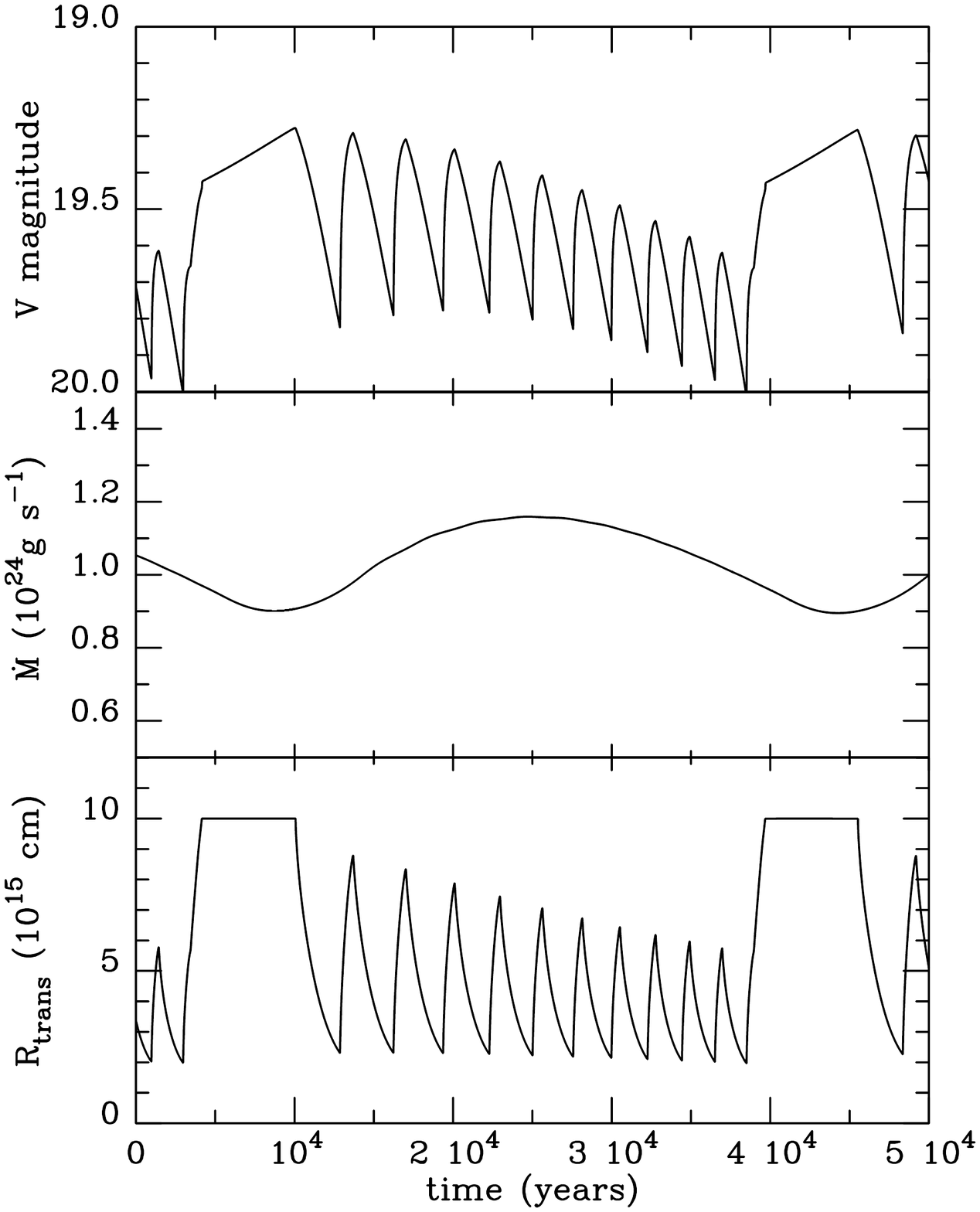}
  \caption{Time evolution of an accretion disc around a 
$10^8\msun$ black hole. The inner and outer radii are at $10^{14}$ and $10^{16}$ cm,
respectively, and the mean mass feeding rate is $10^{24}$g/s ($\dot m \approx 0.01$). Top panel: visual
magnitude; intermediate panel: accretion rate onto the black hole;
lower panel: radius at which the transition between the hot and cold
regimes takes place corresponding to the heating (cooling) front position. [From \cite{HVL09}]}
 \label{fig:sub1}
\end{figure}

\begin{figure}[!]
\centering
  \includegraphics[width=0.6\textwidth]{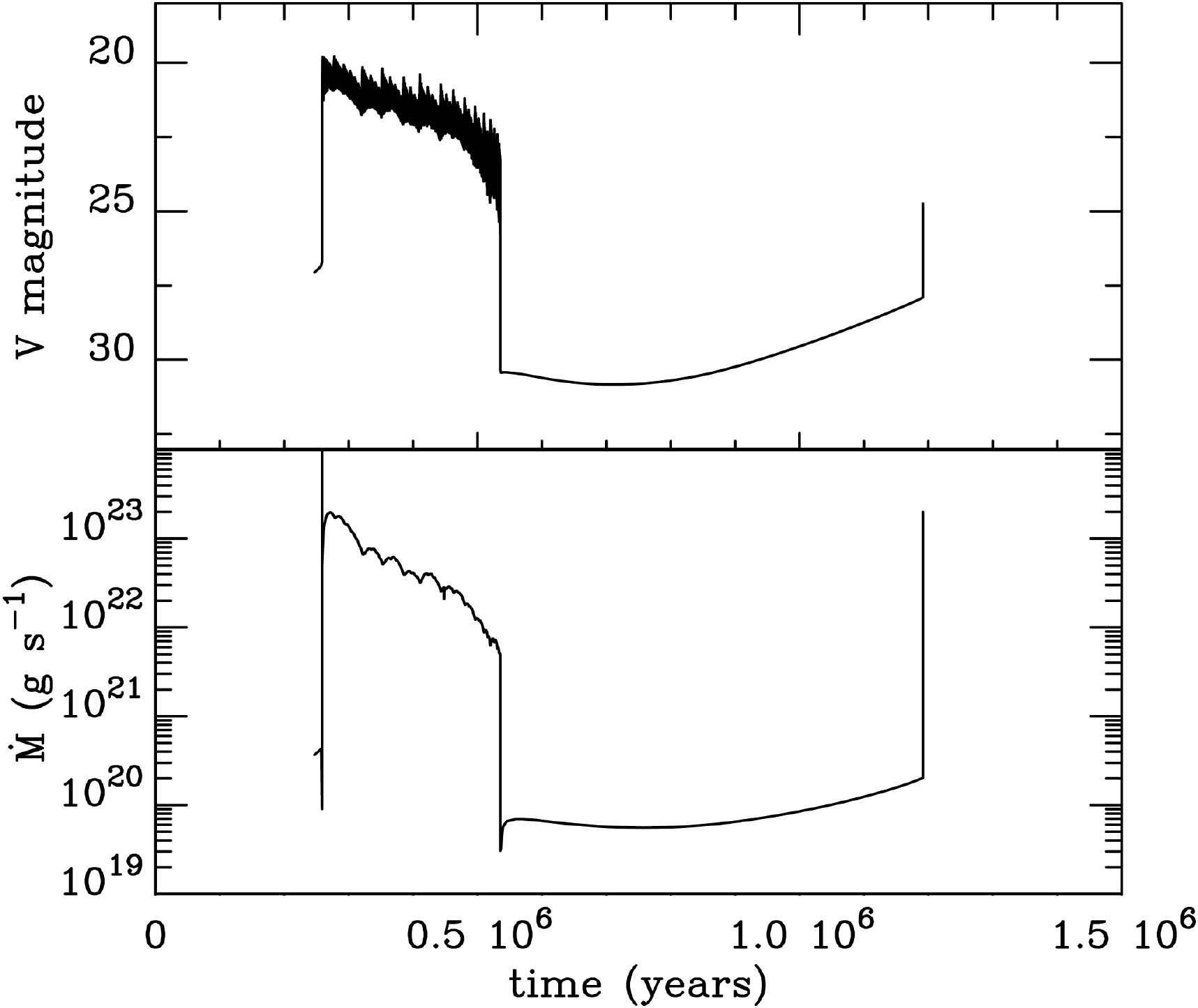}
  \caption{Time evolution of an accretion disc with the same  parameter
  as in Fig. \ref{fig:sub1}, except for the mean mass feeding rate which is here $2\times 10^{22}$g/s ($\dot m \approx 0.0001$). Top panel:
visual magnitude, lower panel: accretion rate onto the black hole.[{\sl From \cite{HVL09}}]}
  \label{fig:sub2}
\end{figure}

As can be seen in Figs. \ref{fig:sub1} and \ref{fig:sub2}, which show the light-curve produced by an unstable disc around a supermassive black hole, the thermal disc instability produces outbursts also  in AGNs,
but a closer look shows that they are rather different from the eruptions observed in systems with less massive compact objects. 

For relatively high mass-feeding rates ($\dot m \approx 0.01$ -- Fig. \ref{fig:sub1}), the instability results in low-amplitude ($\sim 1$ magnitude) outbursts corresponding to tiny modulations of the accretion rate.
In fact, one sees that the timescale of brightness variations, corresponding to the temperature oscillations, is much shorter than the timescale of the accretion--rate variations. 
The basic reason for this is that the
front propagation speed is approximately $\alpha$ times the sound speed
which means that front propagation time-scale is
 \begin{equation}
 t_{\rm front} \approx \frac{R}{\alpha c_{\rm s}} = \frac{R}{H} t_{\rm th},
 \end{equation}
where $t_{\rm th}$ is the thermal time scale. Hence $t_{\rm front}$ is 
shorter than the viscous time $t_{\rm visc} = (R/H)^2 t_{\rm th}$ by
a factor $R/H$, i.e. by several orders of magnitude: 
\begin{align}
\frac{H}{R} & \approx 3.2 \times 10^{-4}T_4^{1/2} m_8^{-1/2}R_{15}^{1/2}\\
&=5.5 \times 10^{-5} T_4^{1/2}\, r^{1/2}.
\label{eq:153}
\end{align}
The viscous time is therefore
\begin{equation}
t_{\rm vis} \approx 1.5 \times 10^4 \alpha^{-1}\,T^{1/2}_4\,m_8\,r^{1/2} \rm \,yr.
\end{equation}

Hence the front propagates so rapidly that the surface density 
does not have time to react to the changes of temperature.

In the case of stellar-mass compact objects, $t_{\rm front}$ is shorter than $t_{\rm visc}$, 
but not by such a large amount and strong gradients in the
disk make the effective viscous time comparable to the front
propagation time. The reason is that while in dwarf-nova stars and X-ray binaries the unstable disc
region is at $r \simeq 10^4$, in AGNs it is rather at $r < 300$, much deeper in the gravitational well
of the accretor.

When one lowers the accretion rate to $\dot m \approx 0.0001$, the disc manages to descend to quiescence (Fig. \ref{fig:sub2}), but 
it does it through a pure viscous decay, without a cooling front propagating down through the whole disc. The outburst phase lasts $3\times 10^5$ yr,
which is roughly the disc viscous time. During this slow decay from maximum, heating/cooling fronts get reflected 400 times, creating as
before low amplitude optical magnitude oscillations and tiny modulations of the accretion rate.

Neither disc irradiation, nor inner-disc truncation change basically the AGN outburst properties. 
This is not surprising since in stellar-mass systems, these effects play an important role 
mainly by affecting the cooling from propagation, which in the AGN case are of no importance.

It remains to be seen if in AGNs, thermal-instability outbursts occur in an observable way and even if they are occurring at all.

\section{Beyond thin discs}
\label{sect:advection}

Until now, we have assumed that accretion discs are thin, i.e. that $H/R \ll 1$.
One of the consequences of this assumption was neglecting the advection terms in the energy and momentum equations
for stationary accretion flows. 
The vertically averaged advective ``cooling'' term can be written as
\begin{equation}
\label{eq:advH}
Q^{\rm adv}=\frac{\Sigma T v_r}{R}\frac{ds}{d\ln  R} = \frac{\dot M}{2 \pi R^2} c_{\rm s}^2 \xi_a,
\end{equation}
where $\xi_a$ is a slowly varying function related to the entropy gradient and is usually $\sim 1$.
One then has
\begin{equation}
\frac{Q^{\rm adv}}{Q^+} \sim \frac{c_s^2}{\Omega_K^2 R^2}\approx \left(\frac{H}{R}\right)^2,
\end{equation}
where we used Eqs. (\ref{eq:qplus}) and (\ref{eq:amintK}). Therefore neglecting the advective term
in the energy equation
\begin{equation}
Q^+= Q^- + Q^{\rm adv}
\end{equation}
is justified if one assumes $H/R \ll 1$.

There are two regimes of parameters where this assumption is
not valid, in both cases for the same reason: low radiative efficiency when the time for radial
motion towards the black hole is shorter than the radiative cooling time. Low density
(low accretion rate), hot, optically thin accretion flows are poor coolers and they are one
of the two configurations were advection instead of radiation is the dominant evacuation-of-energy
(``cooling") mechanism. Such optically thin flows are called ADAFs, for Advection Dominated
Accretion Flows\footnote{Some people prefer to call such flows \textsl{RIAF}s, 
from \textsl{Radiatively Inefficient Flows} but I prefer
to stick to the name \textsl{ADAF}, which I had the pleasure to introduce.}. 
Also advection dominated are high-luminosity flows, accreting at high rates, but
they are called ``slim discs" to account for their property of not being thin, but still being described
as if this were not of much importance.

We shall start with optically thin flows.

\begin{itemize}

\item ADAFs

Advection Dominated Accretion Flows' (ADAFs) is a term describing accretion
of matter with angular momentum, in which radiation efficiency is very
low. In their applications, ADAFs are supposed to describe inflows onto
compact bodies, such as black holes or neutron stars; but very hot,
optically thin flows are bad radiators in general so that, in
principle, ADAFs are possible also in other contexts. Of course in the
vicinity of black holes or neutron stars, the virial (gravitational)
temperature is $T_{\rm vir}\approx 5 \times 10^{12} (R_S/R)$ K,
so that in optically thin plasmas, at such
temperatures, both the coupling between ions and electrons and the
efficiency of radiation processes are rather feeble. In
such a situation, the thermal energy released in the flow by the
viscosity, which drives accretion by removing angular momentum, is not
going to be radiated away, but will be {\sl advected} towards the
compact body. If this compact body is a black hole, the heat will be
lost forever, so that advection, in this case, acts as sort of a `global'
cooling mechanism. There, advection may act only as
a `local' cooling mechanism. (One should keep in mind that, in
general, advection may also be responsible for heating, depending on
the sign of the temperature gradient $--$ in some conditions, near the black hole,
advection heats up electrons in a two-temperature ADAF).

In general the role of advection in an accretion flow depends on the radiation
efficiency which in turns depends on the microscopic state of  matter and
on the absence or presence of a magnetic field. If, for a given
accretion rate, radiative cooling is not efficient, advection is
necessarily dominant, assuming that a stationary solution is possible.\\

\item Slim discs

At high accretion rates, discs around black holes become dominated by radiation
pressure in their inner regions, close to the black hole. At the same time the
opacity is dominated by electron scattering. In such discs $H/R$ is no longer
$\ll 1$. But this means that terms involving the radial velocity are no longer
negligible since $v_r \sim \alpha c_s (H/R)$. In particular, the advective term
in the energy conservation equation $v_r \partial S/\partial R$ (see Eq. \ref{eq:energy})
becomes important and finally, at super-Eddington rates, dominant. When $Q^+=Q^{\rm adv}$
the accretion flow is advection dominated and called a slim disc \citep{Abramowicz0988}.

\item Windy discs

In a super-Eddington accreting disc, the radiative pressure can blow-out matter creating an outflow
that will limit the luminosity to its local Eddington value. In such a case $\dot m \sim r$.

\end{itemize}

\subsection{Advection--dominated--accretion--flow toy models}

One can illustrate fundamental properties of ADAFs and slim discs with a simple toy model.
The advection `cooling'  (per unit surface) term in the energy equation
can be written as
\begin{equation}
Q^{\rm adv} = \frac{\dot M}{2 \pi R^2} c_{\rm s}^2 \xi_a
\label{eq:advterm}
\end{equation}
(Eq. \ref{eq:advH}).

Using the (non-relativistic) hydrostatic equilibrium equation
\begin{equation}
\frac{H}{R} \approx \frac{c_{\rm s}}{v_{\rm K}}
\label{eq:hydroeq}
\end{equation}
one can write the advection term as
\begin{equation}
Q^{\rm adv} = \Upsilon \frac{\kappa_{\rm es}c}{2R}\left(\frac{\dot M}{\eta}\right)\xi_a \left(\frac{H}{R}\right)^2
\label{eq:advterm2}
\end{equation}
whereas the viscous heating term can be written as
\begin{equation}
Q^+= \Upsilon\frac{3}{8}\frac{\kappa_{\rm es}c}{R}\left(\frac{\dot M}{\eta}\right) ,             
\label{visheat}
\end{equation}
where
\begin{equation}
\label{eq:upsilon}
\Upsilon=\left(\frac{c }{\kappa_{\rm es}r}\right)^2 .
\end{equation}
Since $\xi_a\sim 1$, 
\begin{equation}
\label{eq:qadv}
Q^{\rm adv}\approx Q^+\left(\frac{H}{R}\right)^2
\end{equation}
and,
as said before, for geometrically thin discs ($H/R\ll1$) the advective term $Q^{\rm adv}$ is negligible compared
to the heating term $Q^+$ and in thermal equilibrium viscous heating must be compensated by radiative cooling.
Things are different at
very high temperatures, when $(H/R) \sim 1$. Then the advection
term is comparable to the viscous term and cannot be neglected in the equation of thermal equilibrium. In some cases
this term is larger than the radiative cooling term $Q^-$ and (most of) the heat released by viscosity is \textsl{advected} 
toward the accreting body instead of being locally radiated away as happens in geometrically thin discs. 

From Eq. (\ref{eq:amintK}) one can obtain a useful expression for the square of the relative disc height (or aspect ratio):
\begin{equation}
\left(\frac{H}{R}\right)^2 =
              \frac{\sqrt 2}{\kappa_{\rm es}} \left(\frac{\dot m}{\eta}\right)
                     \left(\alpha \Sigma\right)^{-1}r^{-1/2}.
\label{eq:bizarre}
\end{equation}
Deriving Eq. (\ref{eq:bizarre}) we used the viscosity prescription $\nu=(2/3)\alpha c_{\rm
s}^2/\Omega_K$.

Using this equation one can write for the advective cooling
\begin{equation}
\label{eq:advterm3}
Q^{\rm adv} = \Upsilon\Omega_K \xi_a  \left(\alpha \Sigma\right)^{-1}\left(\frac{\dot m}{\eta}\right)^2.
\end{equation}
The thermal equilibrium (energy) equation is 
\begin{equation}
\label{eq:energy_eq}
 Q^+= Q^{\rm adv}+Q^- .
\end{equation}
The form of the radiative cooling term depends on the state of the accreting matter, i.e. on it temperature, density and chemical composition.
Let us consider two cases of accretion flows: 

\begin{description}
\item[$-$] optically thick
 \item{} and 
\item [$-$] optically thin. 
\end{description}
 
For the optically thick case we will use the diffusion approximation formula
\begin{equation}
\label{eq:rad_approx}
Q^-= \frac{8}{3}\frac{\sigma T_c^4}{\kappa_{\rm R}\Sigma},
\end{equation}
and assume $\kappa_{\rm R}=\kappa_{\rm es}$. With the help of Eq. (\ref{eq:bizarre}) this can be brought to the form
\begin{equation}
\label{eq:slimopthick}
Q^-_{\rm thick}=8 \Upsilon \left(\frac{\kappa_{\rm es} R_S}{c}\right)^{1/2}r^2\Omega_K^{3/2} \left(\alpha \Sigma\right)^{-1/2}\left(\frac{\dot M}{\eta}\right)^{1/2} .
\end{equation}
For the optical thin case of bremsstrahlung radiation we have
\begin{equation}
\label{eq:brems}
Q^-=1.24\times 10^{21}H\rho^2T^{1/2}
\end{equation}
which using Eq. (\ref{eq:153}) can be written as
\begin{equation}
\label{eq:opthin}
Q^-_{\rm thin}=3.4\times 10^{-6}\Upsilon r^2\Omega_K \alpha^{-2}\left(\alpha \Sigma\right)^2.
\end{equation}
\begin{itemize}
\item In the \textsc{optically thick} case we have therefore
\begin{eqnarray}
\label{eq:en_thick}
\xi_a \left(\frac{\dot M}{\eta}\right)^{2} +&& 0.18 r^{1/2} \left(\alpha \Sigma\right) \left(\frac{\dot M}{\eta}\right)  +\nonumber \\
&& + 2.3 r^{5/4}\left(\alpha \Sigma\right)^{1/2}\left(\frac{\dot M}{\eta}\right)^{1/2}=0
\end{eqnarray}

\item In the \textsc{optically thin} case the energy equation has the form
\begin{eqnarray}
\label{eq:en_thin}
\xi_a \left(\frac{\dot M}{\eta}\right)^{2} + && 0.18 r^{1/2} \left(\alpha \Sigma\right) \left(\frac{\dot M}{\eta}\right)  + \nonumber \\
&&+ 3\times 10^{-6}\alpha^{-2}r^2\left(\alpha \Sigma\right)^3=0
\end{eqnarray}
\end{itemize}

There are two distinct types of advection dominated accretion flows: optically thin and optically thick.
We will first deal with optically thin flows which are the true \textsl{ADAFs}.
\begin{figure} [h!]
\centering
\includegraphics[width=0.9\textwidth]{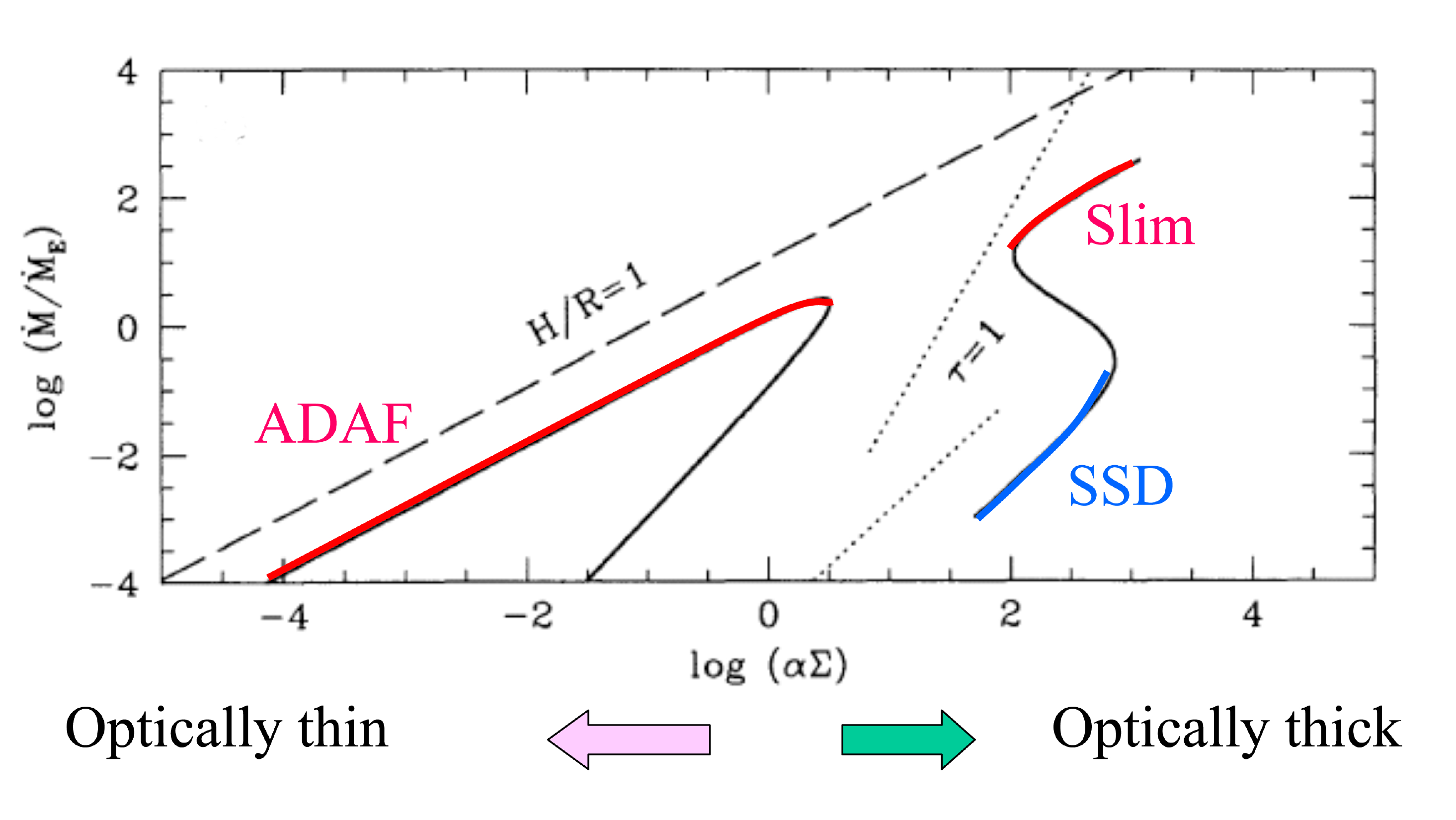}
\caption{Thermal equilibria for optically thick (the right solid S-shaped line) and optically
thin (the left solid line) accretion flows. The upper branches correspond to advection-dominated solution
(ADAF \& Slim). SSD correspond to the Shakura-Sunyaev solution. Flows above the dotted lines $\tau =1 $ are optically thin , $\tau$ being the effective optical
depth calculated for radiation-pressure dominated (upper line) or gas-dominated (lower line) configurations.
This figure corresponds to  $M_{\mathrm{BH}} = 10 \msun$, $r=5$, $\alpha=0.1$ and $\xi_a = 1$. {\sl [Adapted from
\citet{Abramowicz0195}].}}
\label{fig:adaf1}
\end{figure}

\subsubsection{Optically thin flows: \textsl{ADAFs}}
\label{subsub:adaf}
For prescribed values  $\alpha$ and $\xi_a$, Eq. (\ref{eq:en_thin}) is a quadratic equation in $(\dot m/\eta)$ whose solutions in the form of $\dot m(\Sigma)$ describe thermal equilibria at a given value of $R$. Obviously, for a given $\Sigma$ this equation has at most two solutions. 
The solutions form two branches on the $\dot m(\alpha \Sigma)$ -- plane:
\begin{itemize}
\item the ADAF branch
\begin{equation}
\label{eq:adafbranch}
\dot m = 0.53\kappa_{\rm es}\,\eta r^{1/2}\xi_a^{-1} \alpha \Sigma.
\end{equation}
\noindent and\\
\item the radiatively--cooled branch
\begin{equation}
\label{eq:radcoolbranch}
\dot m = 1.9\times 10^{-5}\,\eta r^{3/2}\xi_a^{-1} \alpha^{-2}\left(\alpha\Sigma\right)^2.
\end{equation}
\end{itemize}
From Eqs. (\ref{eq:adafbranch}) and (\ref{eq:radcoolbranch}) it is clear that there exists a maximum accretion rate 
for which only one solution of Eq. (\ref{eq:en_thin}) exists. This implies the existence of a maximum accretion rate at
\begin{equation}
\dot m_{\rm max} \approx 1.7 \times 10^3 \eta \,\alpha^2 r^{1/2}.
\label{eq:mdotmax}
\end{equation}
This is where the two branches formed by thermal equilibrium solutions on 
the $\dot M (\alpha\Sigma) -$ plane meet as seen on Figure \ref{fig:adaf1}.

The value of $\dot m_{\rm max}$ depends on the cooling mechanism in the accretion flow and the
non-relativistic free-free cooling is not a realistic description of
the emission in the vicinity ($(R/R_S)\lesssim 10^3$) of a black hole. The flow there
will most probably form a two-temperature plasma. In such a case
$\dot m_{\rm max}\approx 10 \alpha^2$, with
almost no dependence on radius. For larger radii $\dot m_{\rm max}$
decreases with radius.

\subsubsection{Optically thick flows: \textsl{slim discs}}
\label{subsub:slim}

Since the first two terms in Eq. (\ref{eq:en_thick}) are the same as in (Eq. \ref{eq:en_thin}), the high $\dot m$, advection dominated solution is
the same as in the optically thin case but now represents the
\begin{itemize}
\item Slim disc branch
\begin{equation}
\tag{\ref{eq:adafbranch}}
\dot m = 0.53\, \kappa_{\rm es}\,\eta r^{1/2}\xi_a^{-1} \alpha \Sigma.
\end{equation}

Now, the full equation (\ref{eq:en_thick}) is a cubic equation in $\dot m^{1/2}$ and on the 
$\dot m(\alpha \Sigma)-$ plane its solution forms the
two upper branches of the \bS-curve shown in  Fig. \ref{fig:adaf1}. The uppermost branch corresponds to slim discs while the branch with negative slope 
represents the Shakura-Sunayev solution in the regime \textsl{a.)} (see Sect. \ref{subsec:SS}), i.e. \\
\item a radiatively cooled, radiation-pressure dominated accretion disc
\begin{equation}
\label{eq:radpdom}
\dot m =160\,\kappa^{-1}_{\rm es}\, \eta r^{3/2}\left(\alpha \Sigma\right)^{-1}
\end{equation}
\end{itemize}

\subsubsection{Slim discs and super-Eddington accretion}

From Eqs. (\ref{eq:bizarre}) and (\ref{eq:radpdom}) one obtains for the disc aspect ratio
\begin{equation}
\label{eq:htorradpress}
\frac{H}{R}=0.11\left( \frac{\dot m}{\eta}\right)r^{-1}
\end{equation}
which shows that the height of a radiation dominated disc is constant with radius and proportional to the accretion rate.

But this means that with increasing $\dot m$ advection becomes more and more important (see e.g. Eq. \ref{eq:qadv}) and for
\begin{equation}
\label{eq:adveqplus}
\frac{\dot m}{\eta}\approx 9.2 r
\end{equation}
advection will take over radiation as the dominant cooling mechanism and the solution will represent a slim disc.
Equation (\ref{eq:adveqplus}) can be also interpreted as giving the \textsl{transition radius} between radiatively and
advectively cooled disc for a given accretion rate $\dot m$:
\begin{equation}
\label{eq:advectrans}
{r_{\rm trans}}\approx \frac{0.1}{\eta}{\dot m}
\end{equation}
Another radius of interest is the \textsl{trapping radius} at which the photon diffusion (escape)  time $H\tau/c$ is equal to the
viscous infall time $R/v_r$
\begin{equation}
\label{eq:trapp}
R_{\rm trapp}= \frac{H\tau\,v_r}{c}= \frac{H\kappa \Sigma}{c}\frac{\dot M}{2\pi R \Sigma}=\frac{H}{R}\left(\frac{\dot m}{\eta}\right)R_S.
\end{equation}
Notice that both $R_{\rm trans}$ and $R_{\rm trapp}$ are proportional to the accretion rate.

In an advection dominated disc the aspect ratio $H/R$ is independent of the accretion rate:
\begin{equation}
\label{eq:htorslim}
\frac{H}{R}=0.86\, \xi_a\,r^{1/4},
\end{equation}
therefore contrary to radiatively cooled discs, slim discs do not puff up with increasing accretion rate.

Putting (\ref{eq:htorslim}) into Eq. (\ref{eq:trapp}) one obtains
\begin{equation}
\label{eq:trapp2}
{r_{\rm trapp}}= 0.86\,\xi_a^{-1/2}r^{1/4}\left(\frac{\dot m}{\eta}\right).
\end{equation}
Radiation inside the trapping radius is unable to stop accretion and since $R_{\rm trapp}\sim \dot m$ there is no
limit on the accretion rate onto a black hole.

The luminosity of the toy-model slim disc can be calculated from Eqs. (\ref{eq:slimopthick}) and (\ref{eq:adafbranch})
giving
\begin{equation}
\label{eq:Qminusslim}
Q^-=\sigma T^4_{\rm eff}=\frac{0.1}{\xi_a}\frac{L_{\rm Edd}}{R^2},
\end{equation}
which implies $T_{\rm eff}\sim 1/R^{1/2}$.
The luminosity of the slim--disc part of the accretion flow is then
\begin{equation}
\label{eq:luminslim}
L_{\rm slim}=2\int_{R_{\rm in}}^{R_{\rm trans}}\sigma T^4_{\rm eff}2\pi RdR=\frac{0.8}{\xi_a}L_{\rm Edd}\cdot\ln\frac{R_{\rm trans}}{R_{\rm in}}\approx L_{\rm Edd}\ln\dot m,
\end{equation}
where we used Eq. (\ref{eq:advectrans}).

Therefore the total disk luminosity
\begin{eqnarray}
\label{eq:total}
L_{\rm total}&=& L_{\rm thin} + L_{\rm slim}=\\
&& 4\pi\left(\int_{R_{\rm in}}^{R_{\rm trans}}\sigma T^4_{\rm eff} RdR+ \int_{R_{\rm trans}}^{R_{\infty}}\sigma T^4_{\rm eff}RdR\right)\approx L_{\rm Edd}(1 + \ln\dot m), \nonumber 
\end{eqnarray}
where $L_{\rm thin}$ is the luminosity of the radiation-cooled disc for which Eq. (\ref{eq:teff}) applies.

Such logarithmic luminosity--accretion-rate dependence is observed in the inner regions of simulated accretion flows.

It is easy to see that the same luminosity formula $L\approx L_{\rm Edd}(1 + \ln\dot m)$ is obtained 
when one assumes mass--loss from the disc, resulting in a radially variable accretion rate: $\dot M \sim R$ (see Sect. \ref{sec:disc-jets}).

\section{Disc coronae}

The X-ray emission observed from AGN accretion flows require the presence of emitters other than the optically thick
accretion disc that we have described until now. The highest temperature one can get from the inner regions of such a disc
is $\lesssim 10^6$K (see Eq. \ref{eq:Teff_in}). In analogy to solar and stellar coronae, the optically thin hot structures, much hotter than the underlying
photosphere, the X-ray radiating structure in AGNs and other accreting compact body systems, is called a \textsl{disc corona}.
The main idea is that the hot electrons of such a corona, inverse-Compton upscatter to high energies the soft photons emitted
by the underlying disc.
\begin{figure}[h!]
\begin{center}
\includegraphics[width=0.6\textwidth]{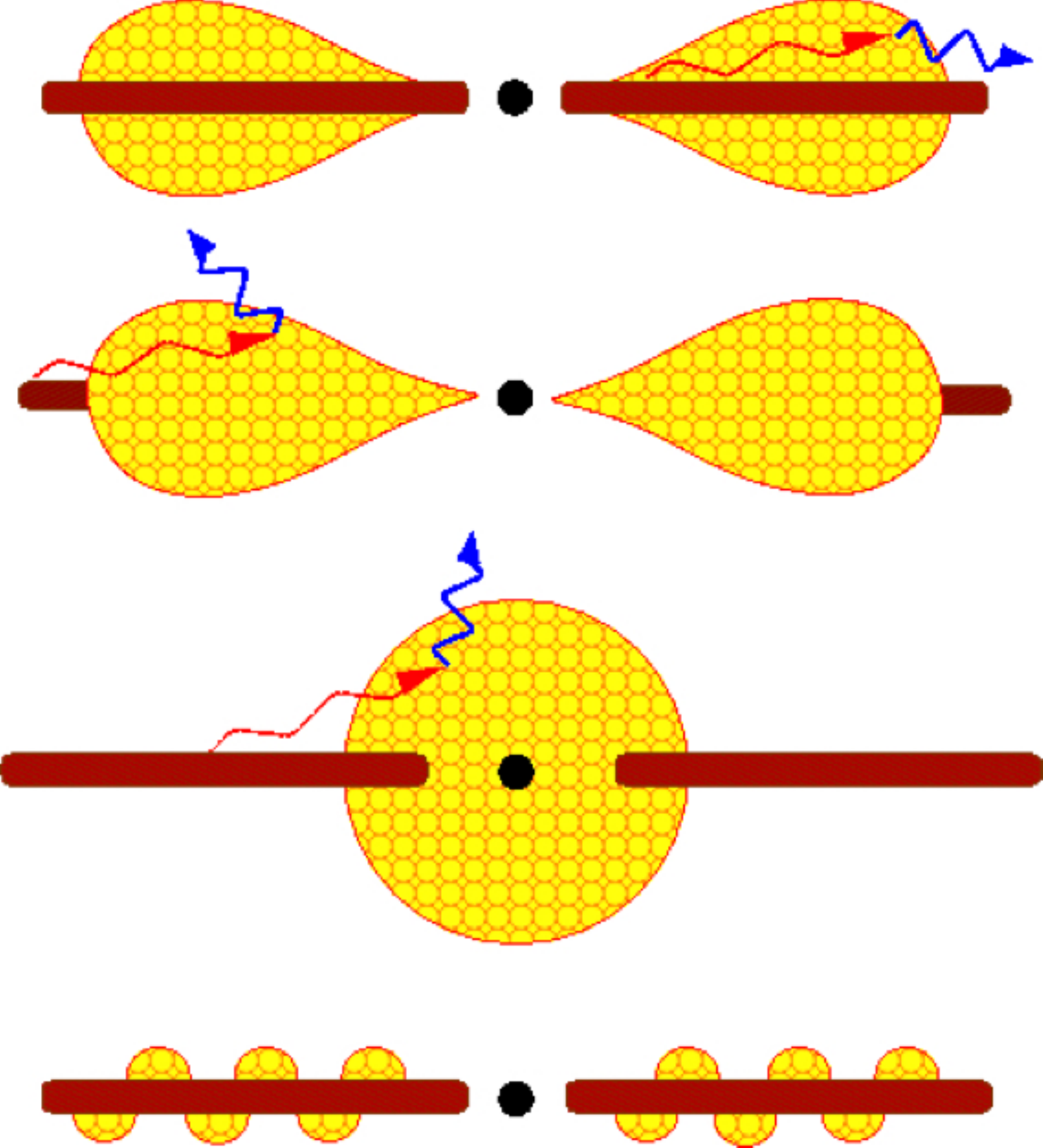}
\caption{Various possible (in principle) disc-coronal structures. [{\sl Courtesy: Chris Done.}]}
\label{fig:corona}
\end{center}
\end{figure}
There is not a single or even leading, or dominating disc corona model. Figure \ref{fig:corona} shows various models or scenarios
that have been proposed in the literature. Various models might correspond to various observed states of the accretion disc. The
bottom configuration corresponds to the oldest proposed model whose idea was that 
differential rotation, together with convection and magnetic fields present in the disc,
could produce loop-like structures, forming a magnetically-confined hot corona,
in analogy to what is observed in the Sun. As for the Sun, magnetic-field reconnection
could play a role in accelerating electrons to high energies. More recently, for different physical reasons
(without invoking convection), models of magnetically heated coronae have been constructed.
In this case it is the MRI amplification mechanism that, from the poloidal field, produces large amount of toroidal field in
the upper layers of the disk.
\begin{figure}[h]
\begin{center}
\includegraphics[width=0.7\textwidth]{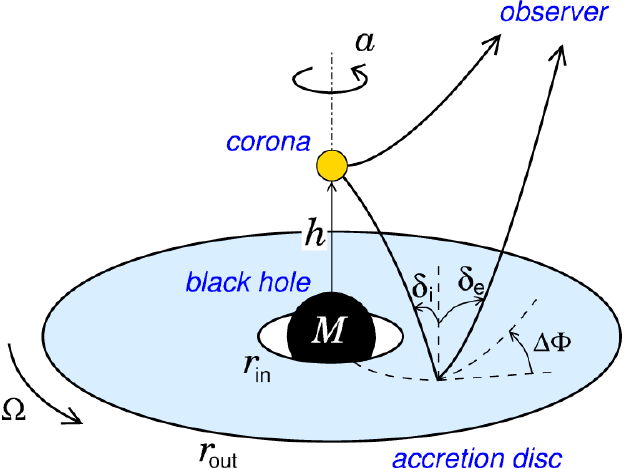}
	\caption{A scheme representing the lamppost ``coronal" model. The feature called ``corona" is the 
	lamppost in question. [{\sl From \cite{Caballero2019}}.]}
\label{fig:lamp}
\end{center}
\end{figure}
Other models invoke accretion-disc ``evaporation",  mechanism which, for low accretion rates, is supposed to produce ADAFs in the inner part of accretion discs.
In fact an ADAF could play the role of a disc corona, as in the second panel from the top.

However, the most used ``coronal" set-up is the so-called \textsl{lamppost} model, whose principle is shown in Figure \ref{fig:lamp}. As it's name demonstrate,
this model does not have exaggerated ambitions to represent a physical reality, but is extremely useful when considering reverberation mapping. Of course this
lamppost is a corona only in name. Most often one says that it could correspond to the base of a jet.

\section{Discs, winds and jets}
\label{sec:disc-jets}

Accretion in AGN is the source of the inverse phenomenon: outflows (ejections) 
that with radiation are the key elements of the feedback cycle
linking the supermassive central black hole to its host galaxy. Radiation and collimated outflows in form of jets
interact with the interstellar medium leading to ejection or heating of the gas, 
but accretion disc winds play apparently
the fundamental role in this interaction. 
Winds and collimated jets are produced by all systems containing a disc-accreting celestial body, from young stellar objects, through white dwarfs and neutron stars, to black holes of all masses,
so one can expect that a common mechanism operates in accretion discs at all scales. The nature of this mechanism is still subject to controversy. 
In the case of winds, three possible mechanisms are invoked. Outer disc regions, most probably eject thermal-driven winds that are accelerated by the thermal gas
pressure. Their velocity is at most $\sim 1000$ km/s.

The radiative pressure due to the intense radiation of most AGN can be a very effective way to drive 
an accretion disc wind. For sub-Eddington accretion rates, UV absorption lines are the main source of opacity (as for winds of massive stars),
while near the Eddington luminosity, Compton scattering is the wind blowing driver. 

\subsection{The ``forgotten" Shakura-Sunyaev solution}

In their seminal 1973 paper, Shakura and Sunyaev considered also the case of super-Eddington accretion and
found a solution, alternative to the slim-disc solution (which they discarded). 
For mass-feeding rates $\dot m > 1$, they identified the \textsl{spherisation} radius where the luminosity is close to the local Eddington value as
\begin{equation}
R_{\mathrm{sph}} \simeq \frac{27}{8} \dot{m} R_{S} \simeq 1 \times 10^{14} \dot{m} m_{8} \,\mathrm{cm}
\end{equation}
Then the solution is obtained by requiring that the local emission within $R_{\mathrm{sph}}$ nowhere exceeds its local
Eddington limit,
This will be true if the outflow is such that the accretion rate through the disc decreases as
\begin{equation}
\dot{m} \simeq \dot{m} \frac{R}{R_{\mathrm{sph}}} .
\end{equation}
Then the total luminosity is
\begin{equation}
L\simeq L_{\rm Edd}(1 + \ln\dot m),
\end{equation}
(see Eq. \ref{eq:total}).
At very high accretion rates ($\dot m \gg 1$) the disc emission will be also strongly beamed by the flow geometry so that an observer situated in the beam of the
emitting system will infer a luminosity
\begin{equation}
\label{eq:beam}
L_{\rm sph}=\frac{1}{b}L_{\rm Edd}(1 + \ln\dot m),
\end{equation}
where $b$ is a beaming factor. Seen from the ``side" such a source with very large apparent luminosities might appear as rather dim.
On the other hand, if e.g., $b\sim 1/\dot m^2$, a stellar mass ultra-luminous X-ray source (by definition $L > 10^{39}$\, erg/s) accreting at
$\dot m \simeq 10^4$, seen along the beam, could have an apparent luminosity $\sim 10^{45}$ erg/s and look like an AGN, but will be positioned
off the galactic center.

\subsection{Relativistic jets}

\begin{figure}[ht]
\begin{center}
\includegraphics[width=\textwidth]{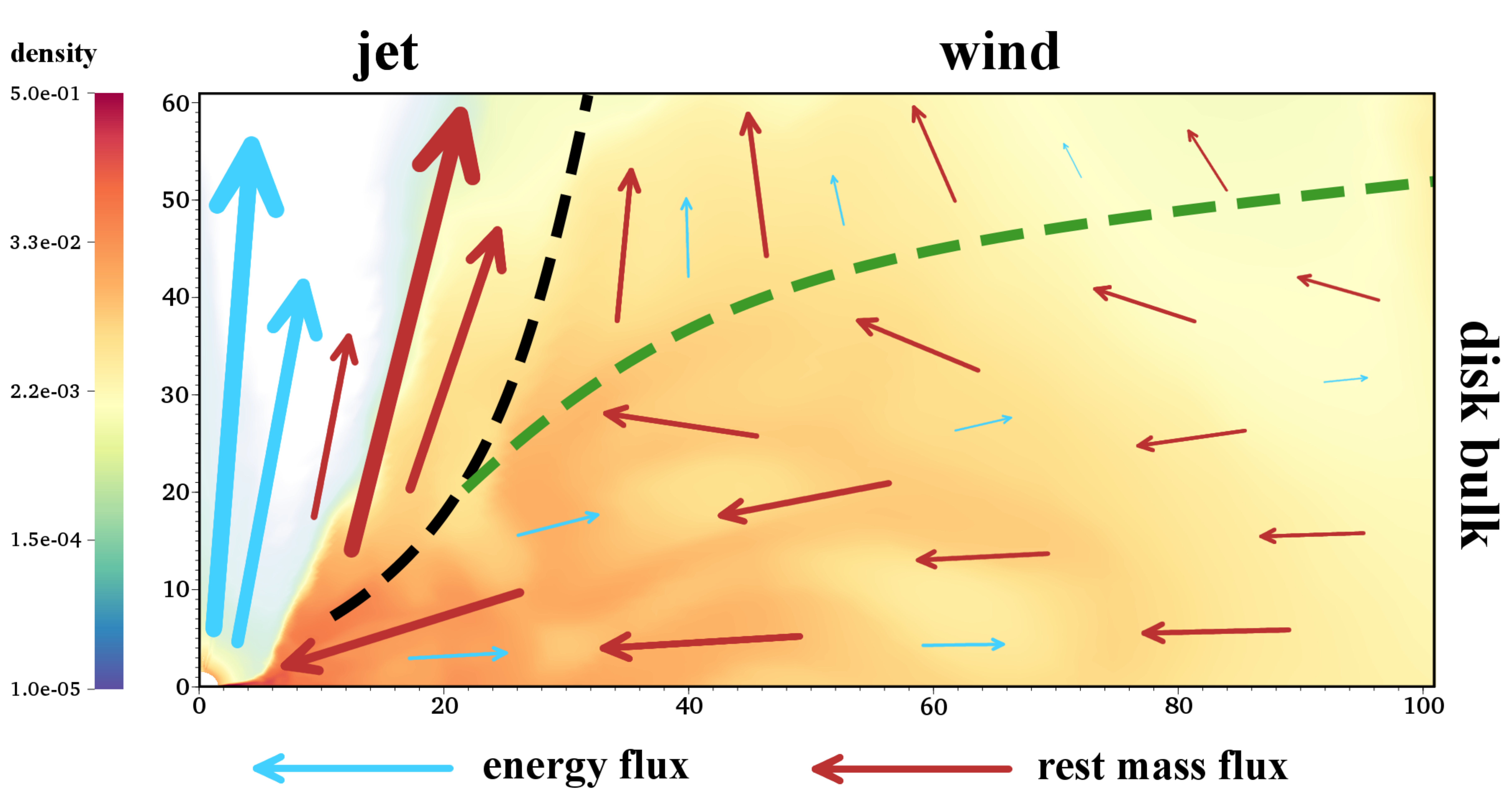}
\caption{Schematic picture of a disc-wind-jet structure near a rotating black-hole. 
	[{\sl From \cite{Sadowski1213}}.]}
\label{fig:jet_wind}
\end{center}
\end{figure}

Magnetic fields may play a fundamental role in the production of disk winds.
In this case the wind is accelerated by the centrifugal force
of the magnetic field lines anchored on the disk and the
magnetic pressure (Blandford-Payne mechanism, cf. \citealp{BP82}).
Magnetic fields 
are usually involved at the generation and acceleration phase and are always supposed to play a decisive role in keeping the jet collimated. Quite often in models, the jet is a collimated
part of an outflow from the accretion disc but in other cases winds and jets can be launched by different mechanisms. Such a case is shown schematically in Fig. \ref{fig:jet_wind}
where the relativistic jet is powered by the black-hole rotation while the source of the sub-relativistic wind energy is gravitational. 

For launching relativistic jets observed in AGN, the best performing models involve large-scale magnetic fields anchored in the
rapidly rotating matter of the inner parts of accretion discs. In the case of a rotating black hole two sources of jet launching energy are possible: the gravitational energy of accretion and the
black-hole rotational energy that can be tapped through the electromagnetic Penrose process. This last mechanism is possible only when the accretor is a black hole because other rotating compact bodies do not have ergoregions.

The Blandford-Znajek (\citealt{BZ77}, hereafter BZ) mechanism, which is the electromagnetic version of the Penrose process works on the same principle as its mechanical analogue:  absorption of negative energy and negative
angular momentum. 
\begin{figure}[h!]
\begin{center}
\includegraphics[width=\textwidth]{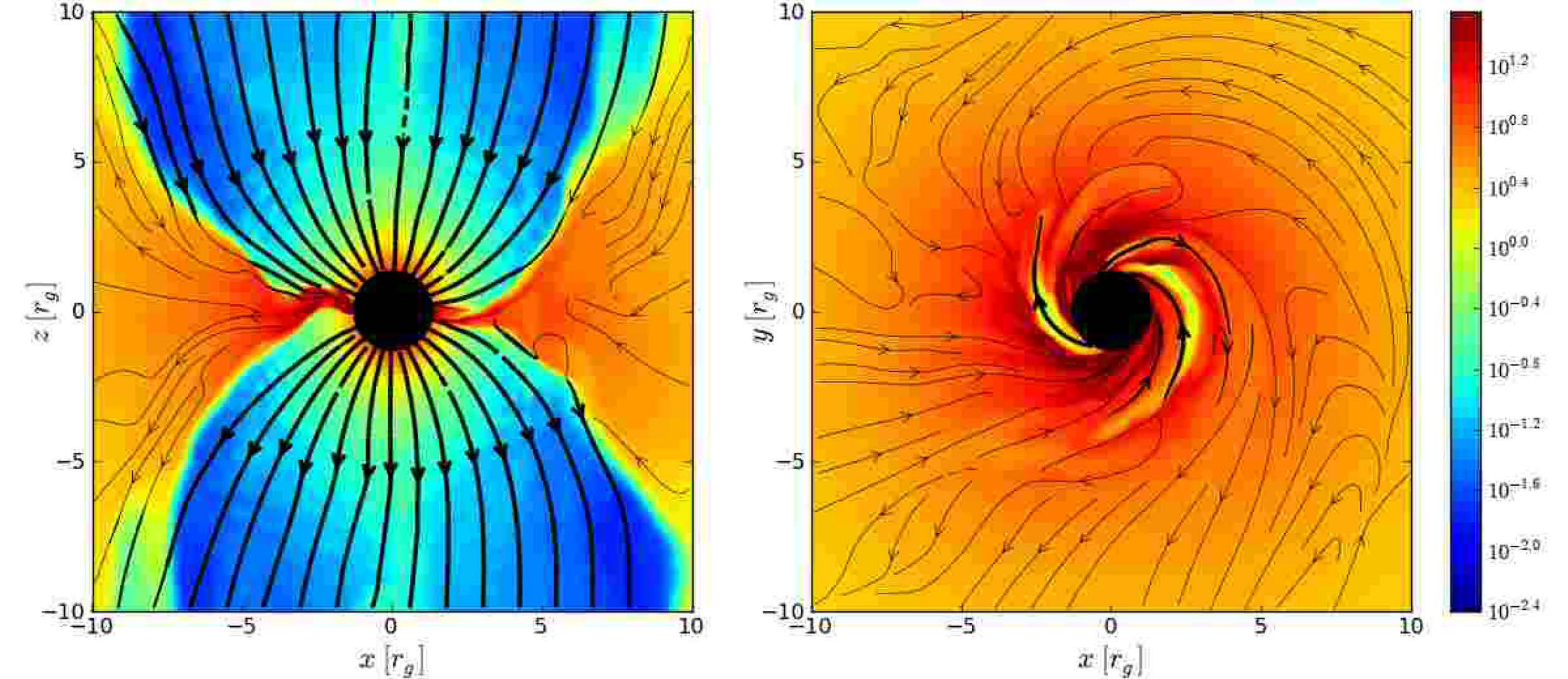}
	\caption{Snapshot at $t \approx 7806\, r/c $ from General Relativistic (GR) MHD simulation of the Blandford-Znajek process  showing logarithm of rest-mass density in colour (see
the scale on the right-hand side) in both the $z-x$ plane at $y = 0$ (top left-hand panel) and the $y-x$ plane at $z = 0$ (top right-hand panel). The black lines trace
field lines, where the thicker black lines show where field is lightly mass-loaded. In this simulation $a=0.9375$ and accreting matter is in form of an ADAF. [{\sl Adapted from \cite{McKinney12}}]}
\label{fig:BZ}
\end{center}
\end{figure}
A typical jet--launching configuration is shown in Fig. \ref{fig:BZ}, where one sees a poloidal magnetic field penetrating the black hole surface being twisted, forming a toroidal component.
The power extracted by the BZ mechanism is equal to
\begin{equation}
\label{eq:BZpower}
P_{\mathrm{BZ}}=\frac{\kappa}{4 \pi c} \Omega_{\mathrm{H}}^{2} \Phi_{\mathrm{BH}}^{2} f\left(\Omega_{\mathrm{H}}\right),
\end{equation}
where
\begin{equation}
\Phi_{\mathrm{BH}}=(1 / 2) \int_{\theta} \int_{\varphi}\left|B^{r}\right| \mathrm{d} A_{\theta \varphi},
\end{equation}
is the magnetic flux threading one hemisphere of the black-hole horizon. ${d} A_{\theta \varphi}$ is the area element in the $\theta - \varphi$ plane.  $\kappa$ is numerical constant depending on the magnetic field geometry ($\kappa = 0.053$ for the so-called split-monopole geometry, such as on Fig. \ref{fig:BZ}. 
Quite often Eq. (\ref{eq:BZpower}) is written with $f(\Omega_H) = 1$, which is a good approximation for black-hole spins up to $a \approx 0.95$, but for larger spins, of special interest when considering the BZ mechanism 
\begin{equation}
f\left(\Omega_{\mathrm{H}}\right) \approx 1+0.345\left(\Omega_{\mathrm{H}} r_{S} / c\right)^{2}- 0.575\left(\Omega_{\mathrm{H}} r_{S} / c\right)^{4}.
\end{equation}
is a more suitable approximation.

One defines the efficiency of the BZ mechanism as the ratio of the time-averaged electromagnetic flux extracting black-hole rotational energy
to the averaged rate at which the black hole absorbs rest-mass energy:
\begin{align}
\eta_{\mathrm{BZ}} & := \frac{\left\langle P_{\mathrm{BZ}}\right\rangle}{\langle\dot{M}\rangle c^{2}} \times 100 \% \\ 
&=\frac{\kappa}{16 \pi }\left(\frac{\Omega_{\mathrm{H}} r_{S}}{c}\right)^{2}\left\langle\phi_{\mathrm{BH}}^{2}\right\rangle f\left(\Omega_{\mathrm{H}}\right) \times 100 \%,
\end{align}
where
\begin{equation}
\label{eq:phi}
\phi_{\mathrm{BH}}={\Phi_{\mathrm{BH}}{}\left(\langle\dot{M}\rangle r_{S}^{2} c\right)^{1 / 2}}\approx 10^4 \eta_{0.1}^{1/2} m_8^{1/2} {\dot m}^{1/2}\left(\frac{\Phi_{\mathrm{BH}}}{0.1\,\rm pc^2 G}\right)
\end{equation}
is the dimensionless magnetic flux
threading the black hole.

Therefore the efficiency of the BZ mechanism depends strongly on the black-hole spin and on the magnetic flux threading the black hole surface. In Eq. (\ref{eq:phi}) we have normalised $\Phi_{\mathrm{BH}}$ by its observed interstellar-medium value. In the most efficient configuration the black hole receives as much large-scale
magnetic flux as can be pushed into it by accretion. By
supplying even more flux than this, some of it remains outside the
horizon where it impedes the accreting gas, leading to a ``magnetically arrested
disc" (MAD; \citealt{Tchekhov1111}). It is the MAD configuration that produces the most powerful relativistic jets.

This requires bringing this field to the innermost disc region where it is needed as an extractor of the black hole rotational energy. It's far from obvious how this can be done.

The simplest problem to address in this context is to consider a poloidal field threading a Keplerian accretion disc. The turbulent-viscosity driven accretion tends to drag the field-lines inwards but due to resistivity they diffuse outwards.
Numerical simulation of this problem have shown that the condition for the magnetic field lines to be significantly dragged inwards is
\begin{equation}
\label{eq:drag1}
 \mathcal{B}=\frac{3 H\left|v_{R}\right|}{2 \eta}=\left(\frac{H}{R}\right) \left(\frac{\nu}{\eta}\right)= \left(\frac{H}{R}\right)\mathcal{P}_{\rm tm} \gtrsim 1,
\end{equation}
where $\nu$, as before, is the kinematic viscosity coefficient, and 
\begin{equation}
\eta=\frac{c^{2}}{ 4 \pi \sigma}
\end{equation}
is the resistivity, with $\sigma$  the electric conductivity. $\mathcal{P}_{\rm tm}$ is known as the magnetic Prandtl number. 
The usual definition of the Prandtl number involves microphysical turbulence and viscosity, but here we are interested
in  transport coefficients of turbulent origin, hence the letter ``t" in the index of the symbol.
Therefore in thin accretion discs  the magnetic field will be dragged inwards only for very large Prandtl numbers:
$\mathcal{P}_{\rm tm} \gg 1$, or $\eta \ll \nu$.  In MRI accretion discs $\mathcal{P}_{\rm tm} \gtrsim 1$, so ``thick" discs, such as slim discs or ADAFs are
needed if the magnetic flux is supposed to accumulate at the inner disc, near the black hole surface \citep{Lubow0394}. 

Another possibility is that the disc's angular momentum is
not removed by turbulent viscosity, but by e.g., magnetic wind. Then Eq. (\ref{eq:drag1}) does not apply and the condition for field dragging becomes
\begin{equation}
\label{eq:drag2}
 \mathcal{B}=\frac{3 H\left|v_{R}\right|}{2 \eta} \gtrsim 1.
\end{equation}
In some models $ \left|v_{R}\right| \sim c_s$ and $\eta \sim Hc_s$, so field dragging is possible.

\bibliographystyle{apalike}
\linespread{0.25}
\renewcommand\bibname{References}
\footnotesize
\bibliography{AGND}

\end{document}